# Magnetization Switching of Single Magnetite Nanoparticles Monitored Optically

S. Adhikari[1]†, Y. Wang[1,2]†, P. Spaeth[1], F. Scalerandi[3], W. Albrecht[3], J. Liu[2], M. Orrit[1]*

† These authors contributed equally to this work

[1] Huygens-Kamerlingh Onnes Laboratory, Leiden University; 2300 RA Leiden, The Netherlands

[2] School of Mechatronics Engineering, Harbin Institute of Technology; Harbin 150001, P. R. China

[3] Department of Sustainable Energy Materials, AMOLF; Science Park 104, 1098 XG Amsterdam, The Netherlands

*Corresponding author. Email: orrit@physics.leidenuniv.nl

**Magnetic nanomaterials record information as fast as picoseconds in computer memories but retain it for millions of years in ancient rocks. This exceedingly broad range of times is covered by hopping over a potential energy barrier through temperature[1], ultrafast optical excitation for demagnetization[2,3] or magnetization manipulation[4], mechanical stress[5,6], or microwaves[7]. As switching depends on nanoparticle size, shape, orientation, and material properties, only single-nanoparticle studies can eliminate ensemble heterogeneity. Here, we push the sensitivity of photothermal magnetic circular dichroism[8] down to *individual* 20-nm magnetite nanoparticles. Single-particle magnetization curves display superparamagnetic to ferromagnetic behaviors, depending on size, shape, and orientation. Some nanoparticles undergo thermally activated switching on time scales of milliseconds to minutes. Surprisingly, the switching barrier appears to vary in time, leading to dynamical heterogeneity, a phenomenon familiar in protein dynamics and supercooled liquids. Our observations will help to identify the external parameters (temperature, magnetic and electric fields, chemical reactions etc.) influencing the switching of magnetic nanoparticles and eventually control switching, an important step for applications in many fields.**

Magnetic nanomaterials, including nanoparticles, promise numerous applications in fields as varied as nanotechnology for data storage, sensing, and logics[9], geomagnetism[10], magnetothermal therapy in medecine[11], or the bio-magnetic compass of bacteria and birds[12]. In all those fields of application, however, the heterogeneity of magnetic nanomaterials is an obstacle to a better characterization and understanding of their magnetic properties. Single-nanoparticles studies[13] are required to overcome ensemble averaging, and open the correlation of magnetic properties with nanoparticle composition, size, shape[14,15], orientation and structure[16]. Several techniques can reach single-nanoparticle magnetization sensitivity, from electrical current measurements[17] to scanning probe microscopies[18–21]. Those techniques, however, are complex and often require contacts and/or scanning probes which may alter the sample's magnetic properties. Non-contact optical techniques are thus particularly attractive. Setting aside X-ray MCD (XMCD) measurements at synchrotrons[15], conventional optical Kerr microscopy based on the magneto-optical Kerr effect (MOKE) lacks the spatial resolution needed to address single nanoparticles, with the notable exception of magnetometry with NV-centers in diamond.[22] We recently proposed an original optical technique, photothermal magnetic circular dichroism (PT MCD) microscopy[8], which has the potential to optically record (time-resolved) magnetic properties of single magnetic nanoparticles (see basic principles of PT MCD in the Supplementary Material). The key advantages of PT MCD over other single-particle methods are that (i) PT MCD is simpler in design and cheaper. It only requires a tabletop microscope in a small-scale lab and, (ii) the sample can be reused after several treatments, providing information about parameters influencing its magnetic properties. In this work, we improved our optical setup and conditions, reaching a detection sensitivity more than 3 orders of magnitude higher than in our previous work. Thereby, we demonstrate the experimental imaging of single 20-nm magnetite nanoparticles and we record their full magnetization curves, one particle at a time. In this method, the single-particle (polar) MOKE signal, which gives rise to a slight magnetic-field-induced difference in optical absorption for right- and left-circularly polarized light, is detected by scattering of a tightly focused probe beam. The resulting magnetization curves hold information about the superparamagnetic or ferromagnetic state of the particles, about their magnetic anisotropy and orientation. In comparison to our previous PT MCD measurements of magnetite nanoparticulate clusters, which were multi-domain clusters of particles, about 400 nm in size, the present measurements are about four orders of magnitude more sensitive and apply to single single-domain particles. In addition, we demonstrate thermally activated switching between two antiparallel magnetization states, and we visualize switching time traces of up to hours with a time resolution as high as ten milliseconds. Magnetic switching, predicted by Néel some 70 years ago, can now be followed by our technique in real time on single magnetite particles, the type of particles which are thought to have recorded paleomagnetic data in ancient rocks. Magnetization curves provide us with estimates of the shape anisotropy and easy-axis orientation of each single nanoparticle according to the Stoner-Wohlfarth model.. A recent article[23] has applied the Stoner-Wohlfarth model in similar way to determine the magnetic anisotropy constant and the easy-axis orientation of 15 single magnetite nanoparticles in a bacterium using XMCD. These particles, however, were larger (~ 50 nm) than ours and, instead of our table-top optical setup, the X-ray microscopy required a synchrotron facility. By varying the applied magnetic field and temperature, we deduce the particles' magnetic dipoles, around $10^5$ Bohr magnetons, and the switching barrier's activation energies, around 0.78 eV. Long switching time traces display pronounced changes in switching rate, i.e., dynamical heterogeneity, indicating that the barrier energy fluctuates significantly with time. Such dynamical heterogeneity is well-known in the dynamics of proteins and of supercooled liquids, had not been reported previously for magnetic nanoparticles.

Figure 1(A) shows a photothermal (PT) image of six single magnetite nanoparticles, labeled P1 to P6, with average diameters ranging from about 19 nm to 25 nm. The sizes, mentioned in the inset in Fig. 1(E-J), are deduced from a comparison of the histogram of photothermal signal of a large number of such nanoparticles (see Fig. S1) to their average diameter, about 19 nm, obtained from transmission electron microscope (TEM) images of 38 particles (see Fig. S2) and assuming a linear relationship between a particle's photothermal signal and its volume (see details about size estimation in the Supplementary Material). A histogram of signal-to-background ratios of 465 single magnetite nanoparticles is shown in Fig. S1, with a mean signal-to-background (S/B) ratio of about 40. Such a high visibility indicates that even smaller magnetite nanoparticles could be detected with our photothermal setup.

By modulating the heating beam between right- and left-handed circular polarizations, we observe the circular dichroism (CD) response of particles P1 to P6, first in the absence of a magnetic field (Fig. 1B). We assign most of the weak signals observed to geometric CD stemming from a low dielectric polarizability of magnetite at our pump wavelength of 532 nm, and from non-mirror-symmetrical (chiral) particle shapes. The significant and consistent positive signal of particle P5, however, suggests a possible ferromagnetic behavior. Upon application of a static magnetic field of $\pm (283 \pm 6)$ mT along the microscope's optical axis, all particles acquire strong CD signals, which change sign with the field direction (Fig. 1C, D), indicating magnetic circular dichroism (MCD). The signal/noise (S/N) ratio exceeds 10 for an integration time of 100 ms/pixel, demonstrating the high sensitivity of the technique (note that we use S/N instead of S/B for the MCD signal because the MCD background fluctuates around zero). The saturation magnetic moment expected for particle P1 can be deduced from its volume and from the side of the cubic unit cell of magnetite, 0.839 nm. With 32 Bohr magnetons per unit cell[24], we expect a magnetic moment of $4.4 \times 10^5$ Bohr magnetons at saturation. The detection sensitivity of our method is thus better than $4.4 \times 10^4$ Bohr magnetons.

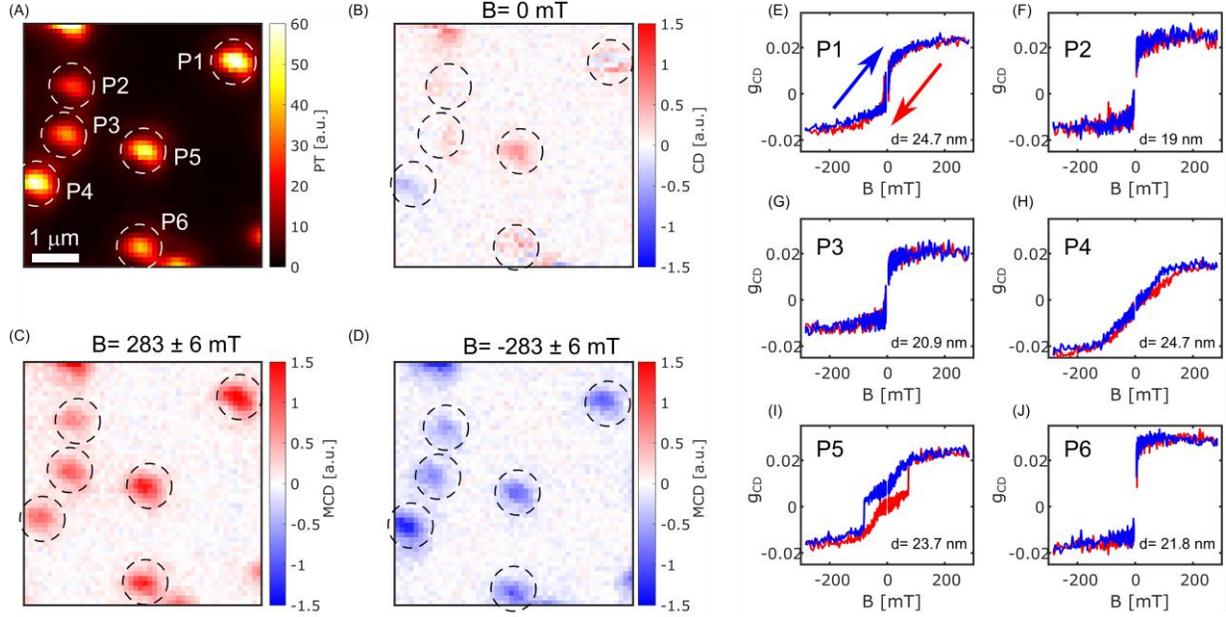

*Fig. 1: Photothermal CD images of single magnetite nanoparticles, about 20 nm in diameter.
**A**: Photothermal (PT) image; **B**: CD image without applied static magnetic field; **C, D**: MCD images with magnetic field applied along the microscope's optical axis (± 283 mT, respectively); the signal units of images (A-D) are mutually consistent; **E-J**: Dependence of the dissymmetry factor $g_{CD}$ on magnetic field for the six nanoparticles P1 to P6. The average diameters of the particles deduced from their PT signal are mentioned in the inset. Particle P1 shows switching between positive and negative dissymmetry factors $g_{CD}$ at weak fields, whereas particle P5 shows hysteresis. The colors indicate the scan direction of the applied field as indicated in (E). Some of these magnetization curves are not (anti-)symmetric around zero field (see nanoparticles P3, P5, P6). We assign these shifts to a weak geometrical chirality of these nanoparticles.*

Our MCD measurements enable us to record the full magnetization curves of single magnetite particles. From the magneto-optical signal $MCD = (I_- - I_+)$, i.e., the difference in circularly left- ($I_-$) and right- ($I_+$) polarized absorption, and from the unpolarized photothermal absorption, $PT = (I_- + I_+)/2$, we deduce the dissymmetry factor, $g_{CD} = 2\frac{I_- - I_+}{I_- + I_+}$. The magnetization curves of particles P1 to P6 in Fig. 1E-J show that the magnetic properties of individual particles are strikingly distinct. According to the Néel-Brown model[25], superparamagnetism is observed when the switching is much faster than the measurement time, whereas ferromagnetism is observed in the opposite limit. Particles P2, P3, P4 and P6 display superparamagnetic behavior, with a regular increase of magnetization with applied field. In contrast, particle P5 shows a typical ferromagnetic behavior indicated by a clear hysteresis loop and a coercive field of about 100 mT. Superparamagnetic particles P2, P3 and P6 show saturation at much lower fields than particle P4. We attribute this difference to the orientation of their magnetic easy axis, which is presumably nearly aligned with the field for particles P2, P3, and P6, but nearly perpendicular to it for particle P4. The magnetization curves can be qualitatively understood and fitted within the Stoner-Wohlfarth model[26] (discussed in more detail in the Supplementary Material, see Fig. S3-S6), which assumes that size and shape anisotropy determine the energy barrier between two opposite magnetization states. From this analysis, we fitted the magnetization curves of P2, P3, P4, and P6 (see Fig. S7) with aspect ratios (1.2, 1.4, 1.8, 1.2) and easy axis angles (50°, 50°, 90°, 40°) with the applied magnetic field, respectively. The magnetization curves of 32 more single particles are presented in Figs. S8, S9 of the Supplementary Material. The low slope of particle P4 is assigned to a high aspect ratio of the particle and to the orientation of its long, easy axis nearly in the sample plane. It is important to note that a ferromagnetic particle with its easy axis perpendicular to the applied magnetic field would show a similar magnetization curve (see details in Supplementary Material, Fig. S22). Particle P1 shows an intermediate behavior, suggesting that magnetic switching might occur during the measurement.

As is apparent from the fluctuations in the MCD signal of particle P1 between positive and negative values (see Fig. 1B), it behaves differently from the other particles which mostly display stable MCD signals. This is confirmed by the large spread of positive and negative $g_{CD}$ for particle P1 at small field values (see image in Fig. 1A and magnetization curve in Fig. 1E). This behavior is absent for particles P2-6. The fluctuations of the MCD signal of particle P1 suggest single-particle magnetization switching. We recorded a 100-s MCD time trace of P1 (Fig. 2A) without an applied field, and indeed find multiple switching events separated by several seconds on average. Time traces of MCD signals of particles P2-6 do not show any magnetization switching (see Fig. S21). We also measured linear dichroism (LD) signals of P1, which do not show any switching (see Fig. S10). As the magnetite particles are mostly about 20 nm in size, we assume that our particles have a single magnetic domain, where exchange energy is minimized by alignment of all spins, producing a macro-spin (see the calculation of the critical radius for a single-domain magnetite nanoparticle in the Supplementary Material). We assign the observed switching events to flips of the macro-spin of particle P1 between two (magnetic-field-dependent) antiparallel states, which we label 'up' and 'down'. From the Néel-Brown theory, we expect switching to be influenced by an applied magnetic field, as we indeed find in our study of the populations of up- and down-levels.

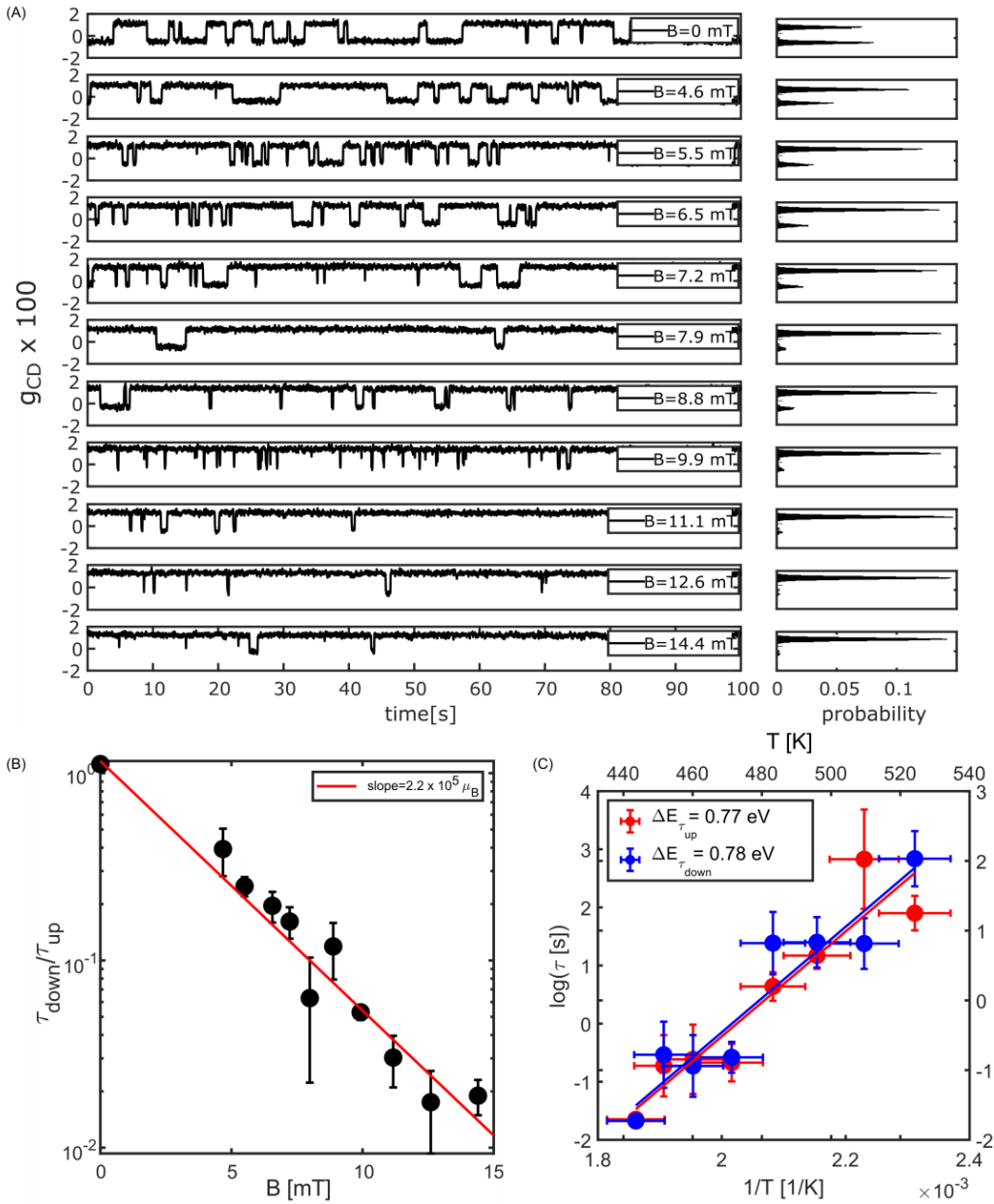

*Fig. 2: Magnetization switching of a single magnetite nanoparticle (particle P1 of Fig. 1) and dependence of the occupation of the two states on magnetic field (B) and on temperature (T). **A**: Magnetization time traces showing switching at different magnetic fields. **B**: Ratio of down- and up-times, $\tau_{down}/\tau_{up}$ versus B with a fit according to the Stoner-Wohlfarth model. The fit slope (inset), $2\mu\cos\psi/k_BT$, provides the particle's magnetic moment. **C**: Temperature dependence of up- and down-times ($\tau_{up}$ and $\tau_{down}$) fitted with a simple Arrhenius law, which provides an energy barrier of $0.78 \pm 0.2$ eV mentioned in the inset, and an attempt frequency of about $10^8$ Hz. The Y error bars are the standard deviations of three measurements at a given temperature. The X error bars are errors in estimation of temperature considering 5% laser power fluctuations.*

Switching time traces of P1 were recorded over 100 s for applied fields varying from 0 to 15 mT, and are presented in Fig. 2A. The population of the up-state (positive $g_{CD}$) increases with the magnetic field. Above 15 mT, mostly the up-state is occupied, as can also be seen in Fig. 1E. A further, weak increase in dissymmetry factor $g_{CD}$, i.e., in magnetization, takes place at higher fields. We assign it to the gradual orientation of the saturated macro-spin along the external magnetic field. Further details of the Stoner-Wohlfarth fit are given in the Supporting Information (SI; Fig. S3). Figure 2A presents histograms of $g_{CD}$ for each time trace, which allow us to perform a change-point analysis and to determine the up- ($\tau_{up}$) and down- ($\tau_{down}$) residence times (see Fig. S11). By fitting the ratio of these times with the Stoner-Wohlfarth model (Fig. 2B), we obtain a slope of $2.05 \times 10^5$ Bohr magnetons. This slope is approximately given by $2\mu \cos\psi/k_B T$, where $\mu$ is the magnetic moment, $\psi$ is the angle between the easy axis and the applied magnetic field, $k_B$ is the Boltzmann's constant and $T$ is the absolute temperature. The angle $\psi$ can be determined from a fit to the magnetization curve, as shown in Fig. S5, and was found to be about $70° \pm 10°$ in the present case. Therefore, we deduce a magnetic moment of the particle of about $3 \times 10^5$ Bohr magnetons, which is reasonably close to the above estimate ($4.4 \times 10^5$ Bohr magnetons). The slight difference between these two values might arise from a magnetic dead layer close to the nanoparticle's surface, i.e., a layer of ill-aligned or disordered spins[27]. A magnetic dead layer with a shell thickness of 1.5 nm would explain the difference between experimental and expected estimates. The switching behavior of two more particles as a function of applied magnetic field is presented in Figs. S12, S13. Note that we neglect the field produced by other particles nearby because of the large separation between neighbor particles required for the optical resolution of single particles (see an estimation of the inter-particle field in the Supporting Information). At short distance from the permanent magnet, i.e., at high field, the magnetic field may not be very uniform. At distances larger than about 1 cm, however, corresponding to fields of some tens of mT, the field is expected to be very homogeneous over the whole field of view and the Hall probe measurement to be reliable (see Fig. S20 and associated discussion).

The Néel-Brown theory[25] of superparamagnetism assigns macro-spin switching to activated barrier crossing, with a rate following an Arrhenius dependence on temperature. To vary the temperature in our measurements, we have varied the laser power of the probe beam, which is tightly focused on the particle under study. Based on literature values of the absorption of magnetite and on COMSOL simulations we estimated the temperature of the single particle P1 to vary from 432 K to 537 K (with an inaccuracy of about 15 K) in the range of probe powers we used (see Supporting Information, Figs. S14, S15). We present the population ratio of up-state and down-state as an Arrhenius plot (Fig. 2C). However, as will be discussed below, the switching rate was found to fluctuate significantly, even at a fixed temperature. We therefore had to average several measurements for each temperature, causing the fairly large error bars on the rates in Fig. 2C. Assuming a simple Arrhenius temperature dependence, i.e., ignoring possible dependences of the attempt frequency and of the barrier energy with temperature, we extract an energy barrier of about 0.78 eV for particle P1 from the slope of Fig. 2C, which is considerably higher than the thermal energy $k_B T$ (0.04 – 0.05 eV). Such a large barrier combined with the exponential Arrhenius dependence explains why switching rates cover many orders of magnitude of times within a comparatively narrow range of temperatures. As the barrier parameters themselves may depend on temperature, however, we stress that our estimate of the energy barrier is only qualitative. Finally, we come back to the aforementioned fluctuations of the barrier rate. Figure 3A shows a magnetization switching trace of particle D3 (see SI) over a duration long enough to observe hundreds of switching events. The switching behavior is obviously much faster in the time interval between 300 and 600 s than at the beginning and end of the trace, although experimental conditions did not change. In this context, it is important to briefly address the stability of the heating laser's output. Laser power fluctuations do not exceed 10% in relative value, and their characteristic times are seconds or less. Therefore, power fluctuations cannot explain the large, sudden, and long-lived rate changes displayed in Fig. 3. Similar rate changes of an activated process are well known in single-molecule traces of complex systems such as enzymes under the concept of dynamical heterogeneity.[28] Histograms of residence times $\tau_{up}$ and $\tau_{down}$ in the two states are shown in Fig. 3B, C. These histograms present a clear excess of events at long times compared to single exponentials, as shown in the insets. As the distribution of switching rates leads to slower-than-exponential decay at long times, we fitted these histograms with stretched exponentials, with stretching exponents of 0.6 for $\tau_{up}$ and $\tau_{down}$. To further prove dynamical heterogeneity, we have used a statistical tool[29] developed earlier for single protein molecules. After coarse-graining the trace by averaging ten consecutive up-times and down-times to reduce statistical fluctuations, we correlate consecutive averaged times for up and down states, separately. The corresponding scatter plots are displayed on logarithmic scales in Fig. 3D, E. A simulation with an exponential distribution of times gives the correlated points displayed as green dots in Fig. 3D, E. The many experimental points falling well outside the green areas confirm that the switching rate itself fluctuates strongly. Similar results are found for another particle monitored over several hours (see Fig. S16). This particle started as super-paramagnetic, then switched to a ferromagnetic behavior, as we verified by measuring a hysteresis loop. After two hours without exposure to laser light, the particle returned to its initial superparamagnetic behavior, and gradually drifted towards ferromagnetism again. After two days in the dark, the particle had returned to a superparamagnetic state.

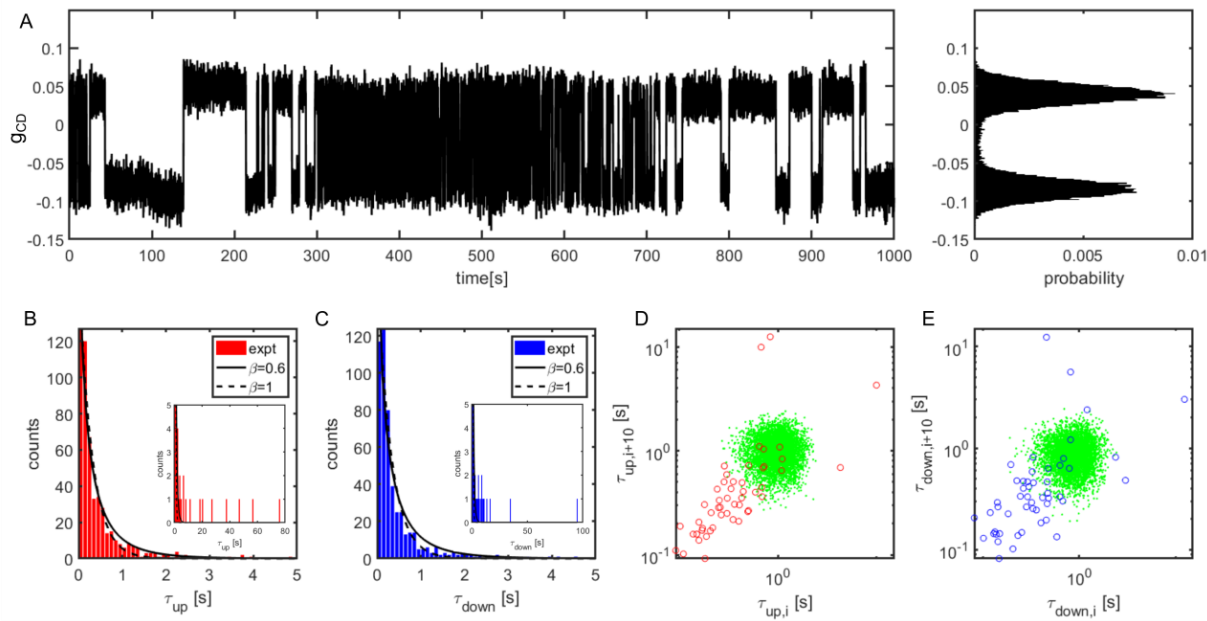

*Fig. 3: Dynamical heterogeneity of magnetization switching of particle D3 from the Supporting Information (see also a similar plot for particle P1 in SI, section 25). **A**: Time trace of magnetization switching over 1000 s. The corresponding histogram of $g_{CD}$ is shown on the right. **B, C**: Histograms of $\tau_{up}$ and $\tau_{down}$ with stretched-exponential fits (stretching exponents $\beta$ given in insets). The insets show a few events with long durations which cannot be properly fitted by (stretched) exponential decays. **D, E**: Empty circles: Correlation plots of successive averages of $\tau_{up}$ and $\tau_{down}$, averaged over ten successive events. The clouds of green dots are obtained by simulation of a single-exponential switching process with the same average time. Deviation of the experimental points from the green cloud highlight the strong dynamical heterogeneity of the trace (see details in the main text). Note the logarithmic scales of times.*

     Dynamical heterogeneity is most often seen as arising from variations of the reaction barrier, through slow conformational changes for proteins[28,29], or through variation of the energy landscape in glassy systems[30], or through changes of the magnetic energy landscape in the case of magnetic nanoparticles[10,15,31]. A transition from ferromagnetism to superparamagnetism has been reported previously[31] for a single iron nanoparticle. However, that study did not report any long time trace with many switching events to support dynamical heterogeneity. In the specific case of magnetite, we speculate that the oxidation state of some iron ions may change through electron transfer or upon oxidation in air, particularly at the elevated temperatures (up to 500 K) caused by laser heating. The resulting changes in the spatial distribution of $Fe^{2+}$ and $Fe^{3+}$ ions, or in the surface binding of ligands by photo- (or temperature-)driven chemistry could change the magnetic energy landscape.[10,16,32] Additional experiments, such as the removal of organic ligands by plasma etching, or ALD coating the particles with 5 nm of $HfO_2$ (see Figs. S17, S18, S19) did not clearly indicate any surface origin of the dynamical heterogeneity. Further experiments are thus needed to explore the role of experimental parameters in switching barrier fluctuations.

In this work, we have imaged and studied individual single-domain magnetite nanoparticles of 20 nm in diameter by purely optical means. This is an improvement of about four orders of magnitude compared to our previous study of single multi-domain magnetite nanoparticulate clusters of 400 nm in diameter. The detection sensitivity reaches about $4\times10^4$ Bohr magnetons. Although our 20 nm magnetite particles had similar sizes, they turned out to be ferromagnetic, super-paramagnetic, or to switch between two antiparallel magnetization states on time scales of milliseconds to minutes. Such information has so far been hidden in ensemble-averaged experiments. The various magnetization curves of single nanoparticles were explained within a simple Stoner-Wohlfarth model, with the anisotropy aspect ratio and the angle of the easy magnetization axis as only fit parameters, adjusted for each particle. The magnetic-field dependence of thermally assisted switching provided us with an estimated magnetic moment of a single magnetite nanoparticle of $10^5$ Bohr magnetons. An anisotropy energy barrier of about 0.78 eV was obtained from the temperature dependence of the switching. The switching rate was found to fluctuate over time, revealing dynamical heterogeneity found earlier in other complex nanometer-scale systems. Such a dynamical heterogeneity commonly observed in protein dynamics or in glassy systems, is new and surprising for purely mineral nanoparticles. Our experiments thus demonstrate the versatility of our technique and the rich information that can be gained at the single-particle level by optical means alone. They open new possibilities to explore the influence of composition, surfaces and defects on nano-magnetic switching, or to study new devices for thermomagnetic actuation, such as antiferromagnetic nanoplatelets[33].

**Methods**

Sample preparation

Magnetite ($Fe_3O_4$) nanoparticles 20 nm in diameter, coated with polyvinylpyrrolidone (PVP) were purchased from Nanocomposix (product number: MGPB20). Hexadecane was purchased from Sigma-Aldrich (product number: H6703). The stock solution of the magnetite nanoparticles was diluted 1,000 times in an aqueous solution. The diluted solution was spin-coated on a UV-plasma-cleaned glass coverslip (thickness about 170 μm) to disperse the particles homogeneously on the glass surface. The glass coverslip was sandwiched with a cavity glass slide with a thickness of about 1.4 mm which contained a cavity to hold the liquid used for the photothermal measurement, hexadecane. All the measurements mentioned in the main text were done on magnetite nanoparticles immersed in hexadecane. In some measurements reported in the supplementary section, magnetite nanoparticles were measured in immersion oil. In that case, the same immersion oil used for the microscope objective was used as the photothermal medium.

Optical setup:

The details of the optical setup are described in our recent publication.[8] Here we give a brief description of the setup and the modifications which have been performed to increase the sensitivity by about 8,000 times. The heating laser (wavelength of 532) nm was passed through an electro-optical modulator (EOM) and a photo-elastic modulator (PEM) which modulated the laser's polarization at frequencies of 33.5 kHz and 50 kHz, respectively. The dual modulation of the laser polarization created a circular dichroism (CD) signal at the sum frequency as discussed in our previous publication[34]. The heating laser was focused in the back-focal plane of the immersion-oil objective (NA= 1.45) and illuminated the sample in a wide-field area of about 3 μm diameter. The collimated circularly polarized continuous-wave probe beam of wavelength 780 nm was focused at the sample using the same objective. The scattered probe beam was detected in the reflection mode, and was filtered from the heating beam using a band-pass filter (BP780) and focused on a photodiode using a lens (focal length 75 mm). The CD signal at the sum frequency, 83.5 kHz was detected using a lock-in amplifier. To vary the magnetic field, a long permanent cylindrical NdFeB magnet (a set of small cylindrical magnets of diameter 3 mm) was placed perpendicular to the sample and its position was varied to change the field. To invert the direction of the magnetic field, the magnetic poles were inverted.

**Data availability**

The data related to the figures and other findings of this study are available from the corresponding author upon reasonable request.

**Code availability**

The code used for the simulation is available from the corresponding author with detailed explanations upon reasonable request.

**Acknowledgments:** We acknowledge the help of Jacqueline A. Labra-Muñoz for the ALD deposition, performed at the Kavli Institute of Nanoscience in Delft University of Technology, financed by NWO through Nanofront project number NF17SYN.

**Funding:** OTP 16008 for PS, Spinoza Orrit for SA, China Scholarship Council for YW

**Author contributions:** SA and MO planned the research, PS realized and adjusted the PT microscope, SA and YW performed the optical measurements, YW did the SW model simulations, FS and WA performed the TEM measurements; all authors contributed in writing and discussing the manuscript.

**Competing interests:** Authors declare that they have no competing interests.

**Supporting Information:**

Supplementary material is available for this paper.

Supplementary Material for

**Magnetization Switching of Single Magnetite Nanoparticles Monitored Optically**

S. Adhikari[1]†, Y. Wang[1,2]†, P. Spaeth[1], F. Scalerandi[3], W. Albrecht[3], J. Liu[2], M. Orrit[1]*

† These authors contributed equally to this work

[1] Huygens-Kamerlingh Onnes Laboratory, Leiden University; 2300 RA Leiden, The Netherlands

[2] School of Mechatronics Engineering, Harbin Institute of Technology; Harbin 150001, P. R. China

[3] Department of Sustainable Energy Materials, AMOLF; Science Park 104, 1098 XG Amsterdam, The Netherlands

*Corresponding author. Email: orrit@physics.leidenuniv.nl

1. Photothermal imaging of single magnetite nanoparticles in hexadecane

To estimate the size of a single magnetite nanoparticle from photothermal measurements, we performed photothermal imaging of many single magnetite nanoparticles as shown in Figure S4. A total of 465 single magnetite nanoparticles were detected and a histogram of cubic root of their photothermal signals is shown in the same figure. Aggregates are distinguished by their very strong photothermal signals (see uncircled spots in Fig. S1) and were not considered in the histogram. The photothermal signal is proportional to a particle's volume and thus, the cubic root of the photothermal signal is proportional to the average diameter of the particle. The histogram of the cubic root of photothermal signals corresponds to the size distribution obtained from the TEM measurements. The peak value of the photothermal histogram corresponds to the peak of value of the size histogram i.e., about 19 nm as obtained from TEM measurements shown in Figure S5. A histogram of signal-to-background ratios of 465 single magnetite nanoparticles is also shown in Figure S4. The peak value is about 40 for 19-nm-diameter particles. Such a high signal-to-background ratio indicates that magnetite nanoparticles smaller than 19 nm can be detected in photothermal imaging.

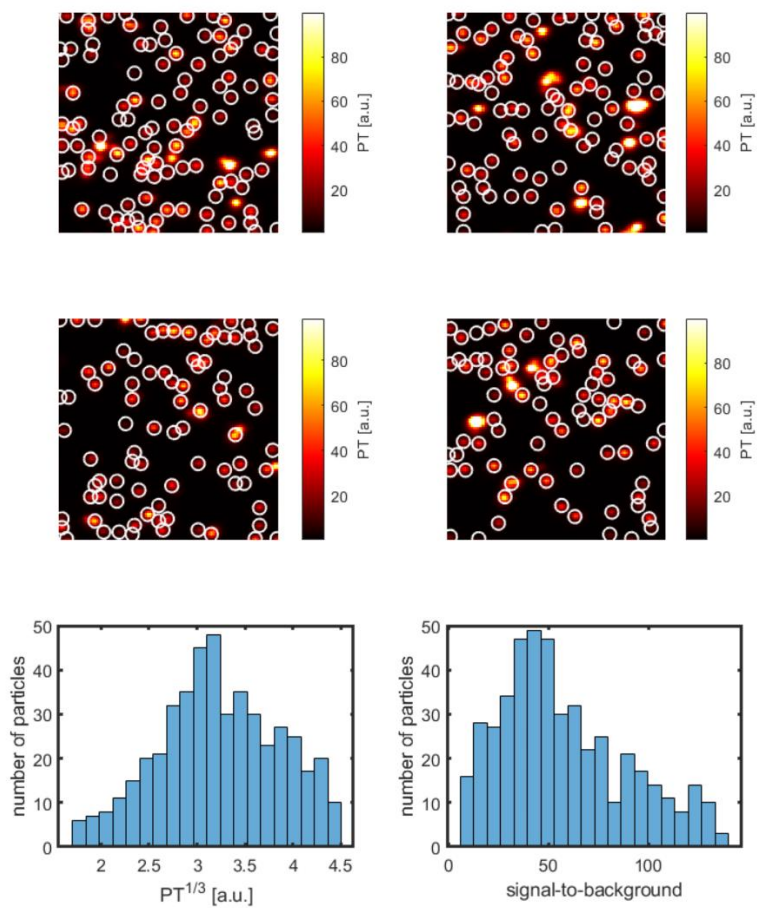

*Figure S4: Photothermal measurements of single magnetite particles in four different regions of the sample. Single particles and aggregates are distinguished by their PT strength. Single particles are marked with white circles. The distribution of the cubic root of maximum photothermal signals of 465 single particles is shown. The peak value of about 3 a. u. corresponds to magnetite particles about 19 nm as shown in Fig. S2. The size mentioned in Figure 1 in the main text is calculated based on this calibration measurement. The cubic root of the PT signal corresponds to the average size of a particle. A histogram of signal-to-background ratios for the 465 single particles is also shown.*

2. TEM images of single magnetite nanoparticles

To get an idea of the size and shape of single magnetite nanoparticles and of their crystallinity, we performed transmission electron microscopy (TEM) imaging as shown in Figure S5. From the TEM images, it is evident that most particles are single-crystalline and heterogeneous in their sizes and shapes. Most particles have asymmetric shapes. To calculate the size of each particle, it is assimilated to a prolate ellipsoid and its volume is calculated. From the ellipsoid volume, the effective average diameter of the particle is calculated as that of the sphere with the same volume. The average diameter obtained from 38 single particles is 19 nm ± 0.5 nm, i.e., close to the size of about 20 nm provided by the manufacturer.

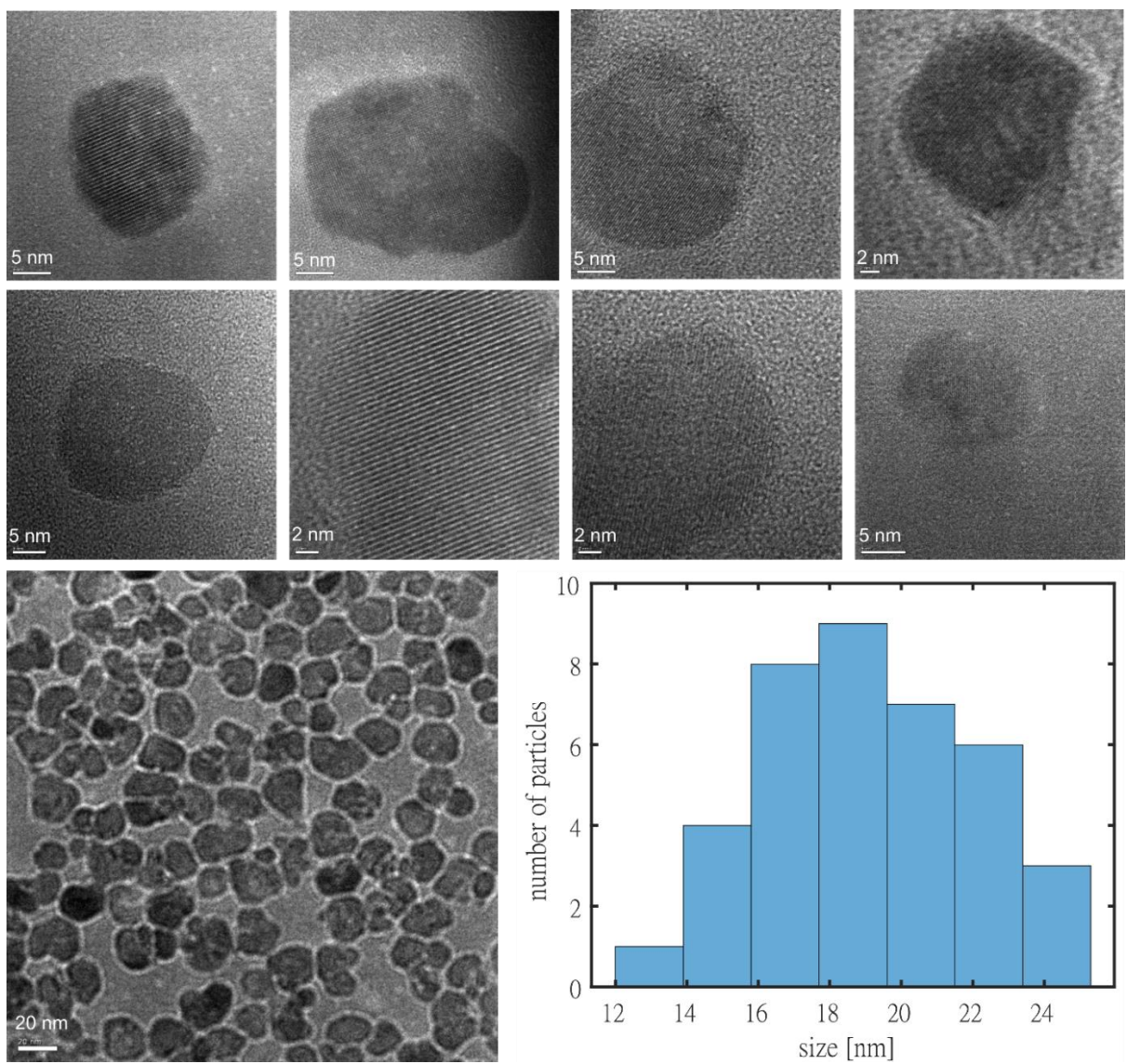

*Figure S5: The top two rows show TEM images of eight single-crystalline magnetite nanoparticles. (Bottom left) TEM image of many single magnetite nanoparticles showing heterogeneity in size and shape. The image was taken slightly out-of-focus for easier identification of particle boundaries. (Bottom right) Histogram of effective diameters of single particles determined from the TEM data.*

## 3. Stoner-Wohlfarth model

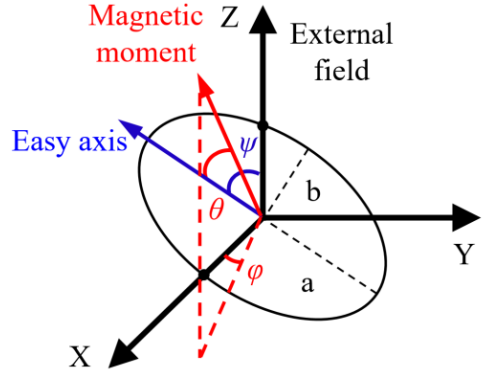

*Figure S6: Schematic representation of a prolate ellipsoid with the angles used for the Stoner-Wohlfarth model. θ is the angle between the easy axis and the magnetic moment, ψ is the angle between the external magnetic field and the easy axis and φ is the azimuthal angle. a and b are the lengths of the semi-major and semi-minor axes. The aspect ratio is a/b.*

Considering a magnetite nanoparticle as a prolate spheroid (Figure S6), the energy of a magnetic state can be calculated using the Stoner-Wohlfarth model as:

$$E = K_{eff} V \sin^2\theta - \mu B(\cos\theta\cos\psi + \sin\theta\sin\psi\cos\varphi),$$

where $V$ is the volume of the particle, $\theta$ is the angle between the easy axis and the magnetic moment and $\varphi$ is the azimuth angle of the magnetic moment $\mu$ ($\mu = M_s V$, where $M_s$ is the saturation magnetization), $B$ is the applied external field, $\psi$ is the angle between the easy axis and the external magnetic field. The effective anisotropy constant $K_{eff}$ can be written as $K_{eff} = K_1 + \frac{\mu_0 M_s^2 (N_x - N_z)}{2}$, where $\mu_0$ is the vacuum permeability, and $K_1$ possibly accounts for anisotropy contributions not arising from the shape (such as the magneto-crystalline and surface anisotropy), which we neglect hereafter ($K_1 = 0$). The second part is the shape anisotropy which we assume to dominate in our case. $N_x$ and $N_z$ are the demagnetization factors of the nanoparticle. For spheroidal nanoparticles, the shape anisotropy is determined by the aspect ratio $k = a/b$, yielding the following demagnetization factors for the cases of oblate, sphere, and prolate spheroids[1]:

$$N_z = \frac{1}{1-k^2}\left[1 - \frac{k}{\sqrt{1-k^2}}\arccos(k)\right], k < 1$$

$$N_z = \frac{1}{3}, \quad k = 1$$

$$N_z = \frac{1}{k^2 - 1}\left[\frac{k}{\sqrt{k^2 - 1}}\operatorname{arccosh}(k) - 1\right], k > 1$$

$$N_x = (1 - N_z)/2$$

Hereafter, we model our nanoparticles as prolate spheroids, as the case of oblate spheroids presenting two degenerate axes is marginal. Under thermal equilibrium conditions ($\tau < t_m$ where $\tau$ is the residence time in a certain magnetic state and $t_m$ is the measurement time), the average magnetization with a given orientation ($\psi$) of the easy axis with respect to the external magnetic field direction is given by:[2]

$$m(\psi) = \frac{M}{M_s} = <\cos(\theta - \psi)> = \frac{\iint (\cos(\theta\text{-}\psi)\cdot \exp\left(-\frac{E}{k_BT}\right)\sin\theta d\theta d\varphi}{\iint \exp\left(-\frac{E}{k_BT}\right)\sin\theta d\theta d\varphi},$$

where $M_s$ is the saturation magnetization, $k_B$ is the Boltzmann constant and $T$ is the absolute temperature. We can get average magnetization curves for different aspect ratios and easy axis orientations using the above equation. Figure 1 of the main text shows different cases of magnetization curves. The case of particle P6 shows a quick saturation of the magnetization for a weak field, with nearly no further increase when the field is further raised. We assign this curve to a particle with its easy axis nearly collinear with the applied field, leading to an early saturation. Particles P2, P3, and to a lesser extent P1, show a steep initial magnetization increase for a weak field, followed by a more gradual increase for higher fields. We assign this behavior to particles with tilted easy axes, which first saturate their magnetic moment along the easy axis, then rotate their magnetization axis upon competition of the anisotropy with the Zeeman energy. In that case, a first saturation along the easy axis gives a component $M_s \cos\psi$ along the applied field, whereas a further increase of the field moves the magnetization towards the applied field, ending up with projection $M_s$ on the applied field. Finally, the case of P4 is assigned to a particle whose easy axis is nearly perpendicular to the applied field. No first saturation can be distinguished, and the magnetization gradually rotates from horizontal to vertical along the applied field. For aspect ratios close to 1, the nearly spherical particle magnetizes along the applied field, producing a magnetization curve very similar to that of P6. These behaviors are exactly obtained in the Stoner-Wohlfarth simulations presented in Figure S7, so that angle and aspect ratio can often be determined by visual comparison of the measured magnetization curves to the simulations.

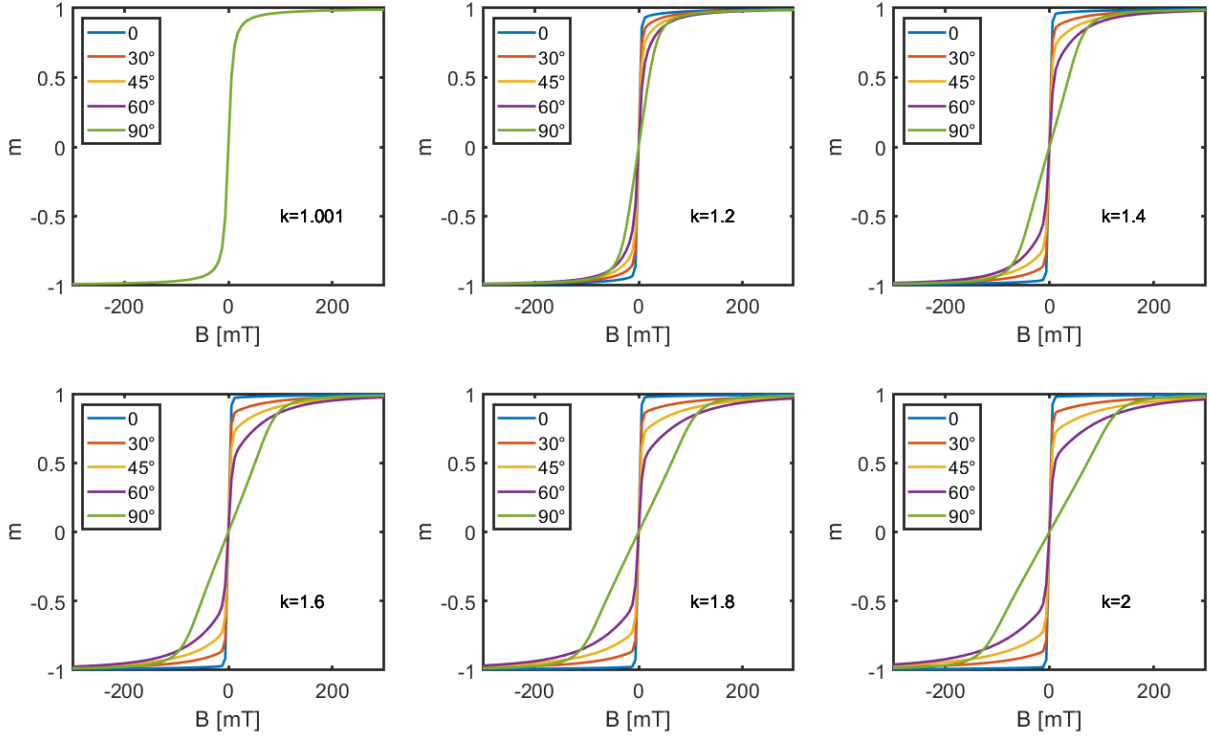

*Figure S7: Magnetization curves simulated using the Stoner-Wohlfarth model for particles with different aspect ratios (k) and angles (ψ) of the easy axis with respect to the external magnetic field.*

If the temperature is too low to establish thermal equilibrium on time scales much shorter than the measurement, i.e., if the thermal energy $k_BT$ allows occasional crossing only of the energy barrier during the measurement, the magnetization will be seen to switch between two states. Then, the population ratio of the two states i.e., the ratio of residence times in the two states follows from Boltzmann statistics as:

$$\frac{\tau_1}{\tau_2} = \exp\left(\frac{E_2 - E_1}{k_BT}\right).$$

Now, using the energy from the Stoner-Wohlfarth model, we can write this ratio for a small enough applied field (i.e., assuming that the applied field is too weak to change the magnetic moment's direction) as:

$$\log\left(\frac{\tau_1}{\tau_2}\right) = \frac{2\mu\cos\psi}{k_BT} B.$$

It turns out that the plots of $\log\left(\frac{\tau_1}{\tau_2}\right)$ vs. $B$ are very close to straight lines in all cases. Therefore, if we plot $\log\left(\frac{\tau_1}{\tau_2}\right)$ vs. $B$, we can calculate the magnetic moment ($\mu_{fit}$) as

$$\mu_{\text{fit}} = \frac{k_B T}{2\cos\psi} \frac{d}{dB}\left(\log\left(\frac{\tau_1}{\tau_2}\right)\right)$$

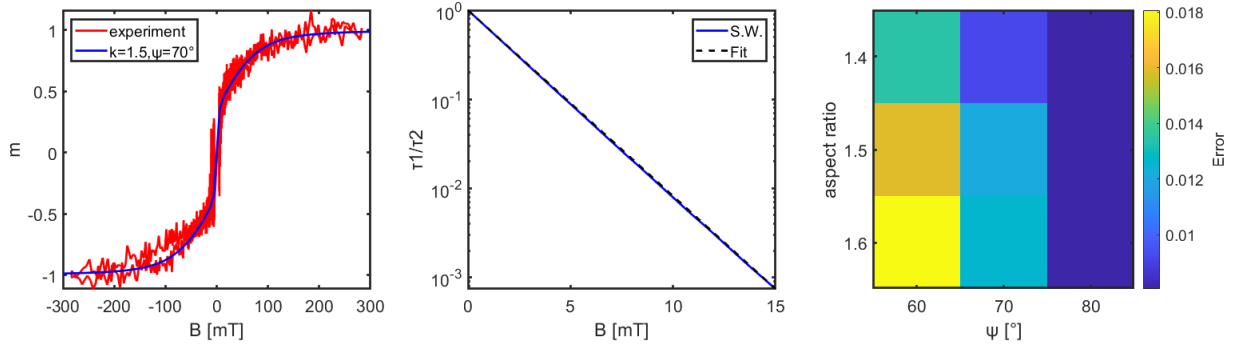

*Figure S8: (Left) a magnetization curve fitted with the Stoner-Wohlfarth model. The aspect ratio (k) and the angle of the easy axis (ψ) are mentioned in the inset. (Middle) The plot* $\log\left(\frac{\tau_1}{\tau_2}\right)$ *vs. the external magnetic field (B) with a linear fit and with the fit using the Stoner-Wohlfarth model. (Right) The error map for the linear fit and the Stoner-Wohlfarth fit for different values of k and ψ.*

Figure S9 shows the energy landscape in the XZ plane under an external magnetic field. When there is zero or weak magnetic field, the magnetic moment prefers aligning to the minimum energy direction i.e. along the easy axis. There are two minimum energy points corresponding to up and down spin states as schematically shown in Figure S9. With the increase in magnetic field strength, the energy diagram shifts towards one minimum and the magnetic moment gradually aligns to the magnetic field direction. The rate of the change in the energy diagram with increase in magnetic field depends on the orientation of the magnetic field with respect to the easy axis, i.e., on angle ψ.

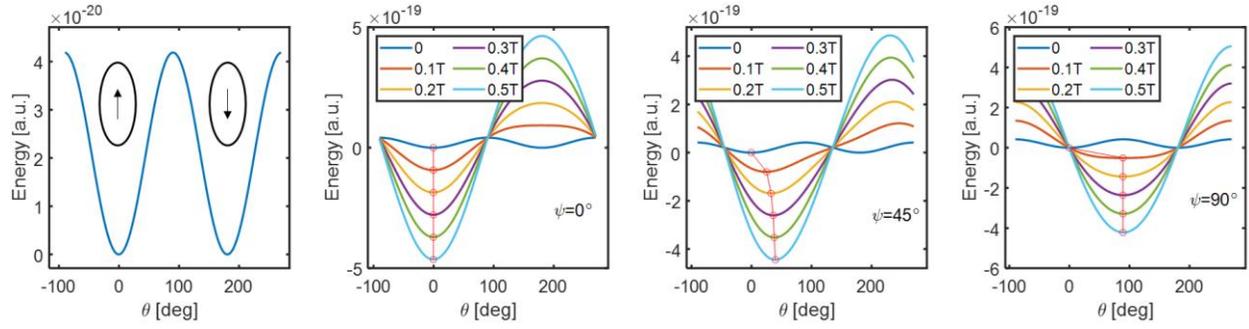

*Figure S9: the energy landscape in XZ plane of a single magnetite nanoparticle. (Left) the energy landscape when there is no external magnetic field. (Middle and right) The energy diagram at different field strengths at three different angles of the field orientation w.r.t the easy axis. The red lines with circle show the minimum energy points.*

4. Magnetization curves of particles P2, P3, P4 and P6 with Stoner-Wohlfarth fits

Figure S10 shows magnetization curves of particles P2, P3, P4 and P6 as labelled in Figure 1 in the main text. The magnetization curves are fitted with the Stoner-Wohlfarth model considering different aspect ratios ($k$) and angles between the applied magnetic field and the easy axis ($\psi$). The optimized values are obtained by minimizing the error of the fit to the experimental data and shown in the insets of center panels in Figure S10.

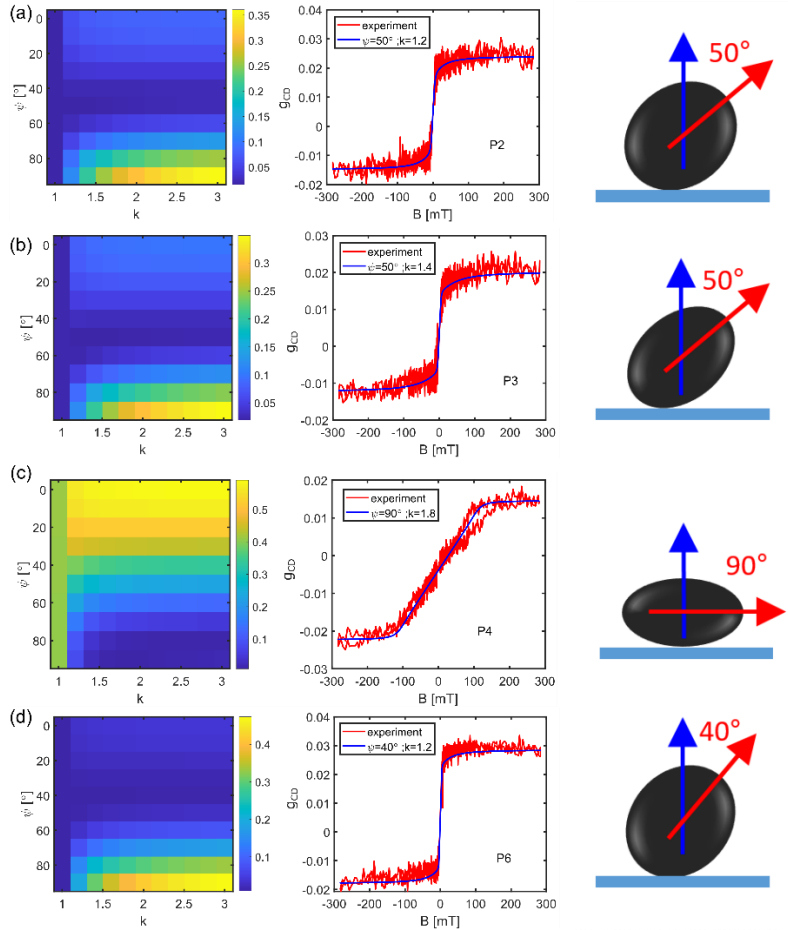

*Figure S10: Magnetization curves of four particles (P2, P3, P4 and P6 as labelled in Figure 1 in the main text) with Stoner-Wohlfarth fits. (Left) An error map of the fit from the experimental data at different aspect ratios (k) and angles between the easy axis and the applied field direction (ψ). (Center) The obtained k and ψ values obtained for the minimum error are shown in the insets and the corresponding Stoner-Wohlfarth fit (blue) is plotted along with the experimental data (red). Particle labels are mentioned. (Right) Schematic representation of the magnetic moment (red) along the prolate particle axis for low enough applied magnetic field (blue).*

## 5. Magnetization curves of 32 single magnetite particles

We measured the magnetization curves of 32 single magnetite nanoparticles as shown in Figure S11 and Figure S12. We categorized them in three types, Type I, Type II and Type III. Type I particles are superparamagnetic, Type II particles present intermediate behaviors, i.e., magnetization switching and Type III particles are ferro(i)magnetic. Magnetite is ferrimagnetic, but we will use the usual terminology of superparamagnetism and refer to its magnetized ferrimagnetic state as 'ferromagnetic' in the following. The particles that have magnetization curve similar to particle P4 shown in Fig. S7, have in-plane easy axis. It is difficult to distinguish whether these particles are superparamagnetic or ferromagnetic as ferromagnetic particles with in-plane easy axis would also show similar magnetization curves (see later in Fig. S22). These kinds of particles in Fig. S8 and Fig. S9 are termed Type I or Type III.

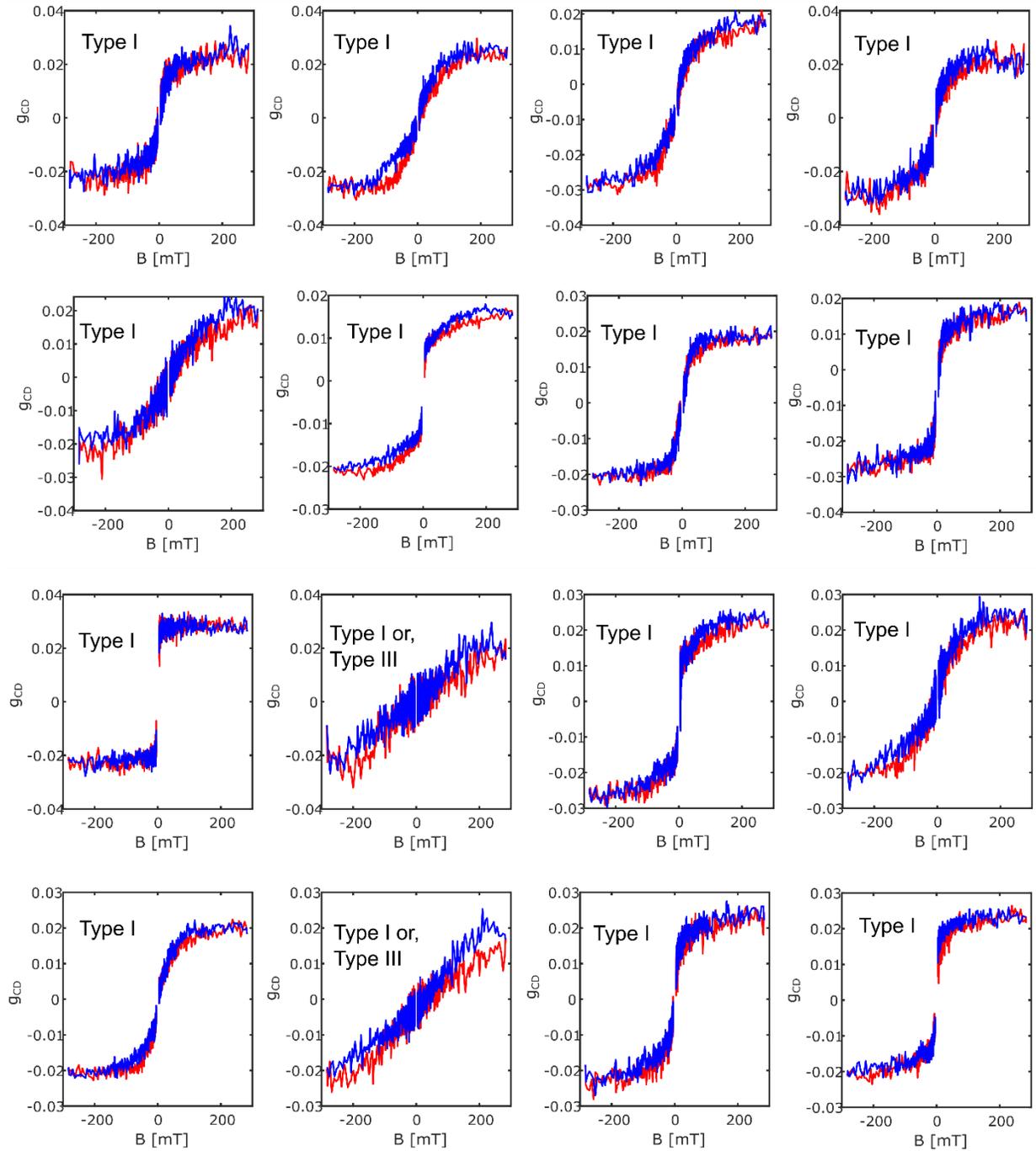

*Figure S11: Magnetization curves of 16 single magnetite nanoparticles of Type I and Type III. Red: scan from positive to negative applied magnetic field, blue: scan from negative to positive applied magnetic field.*

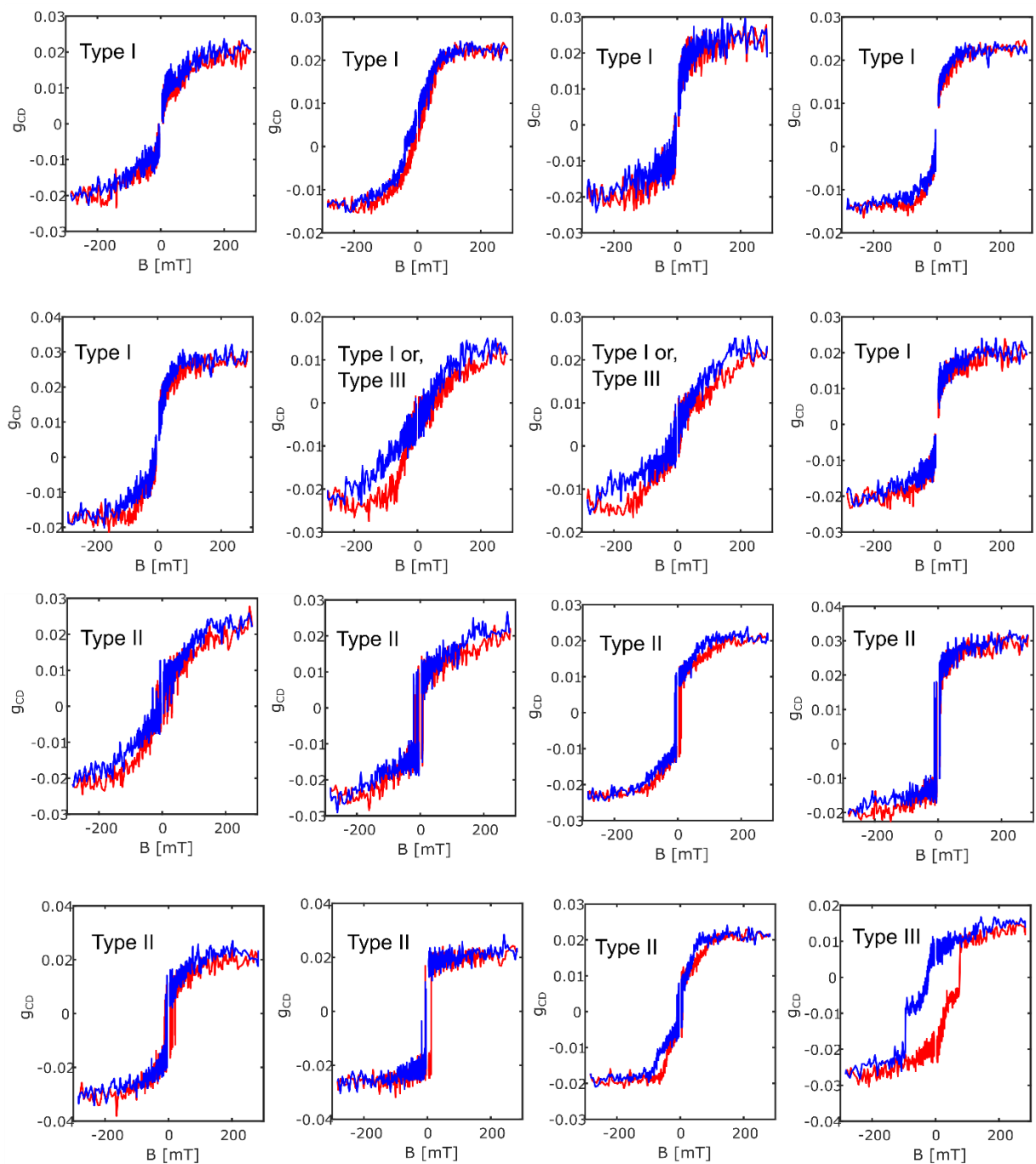

*Figure S12: Magnetization curves of 16 single magnetite nanoparticles of three different types, Type I, Type II and Type III. Red: scan from positive to negative applied magnetic field, blue: scan from negative to positive applied magnetic field.*

## 6. Time traces of PT, CD, LD and LD45 signals for particle P1

Figure S13 shows time traces of photothermal absorption (PT), circular dichroism (CD), and linear dichroism at two different polarization orientations, at 0°/90° (LD) and at 45°/-45° (LD45). Although there are some signal fluctuations for the case of PT, LD and LD45, none of them show signal fluctuations between two states. The only trace showing clear switching between two states is the CD time trace.

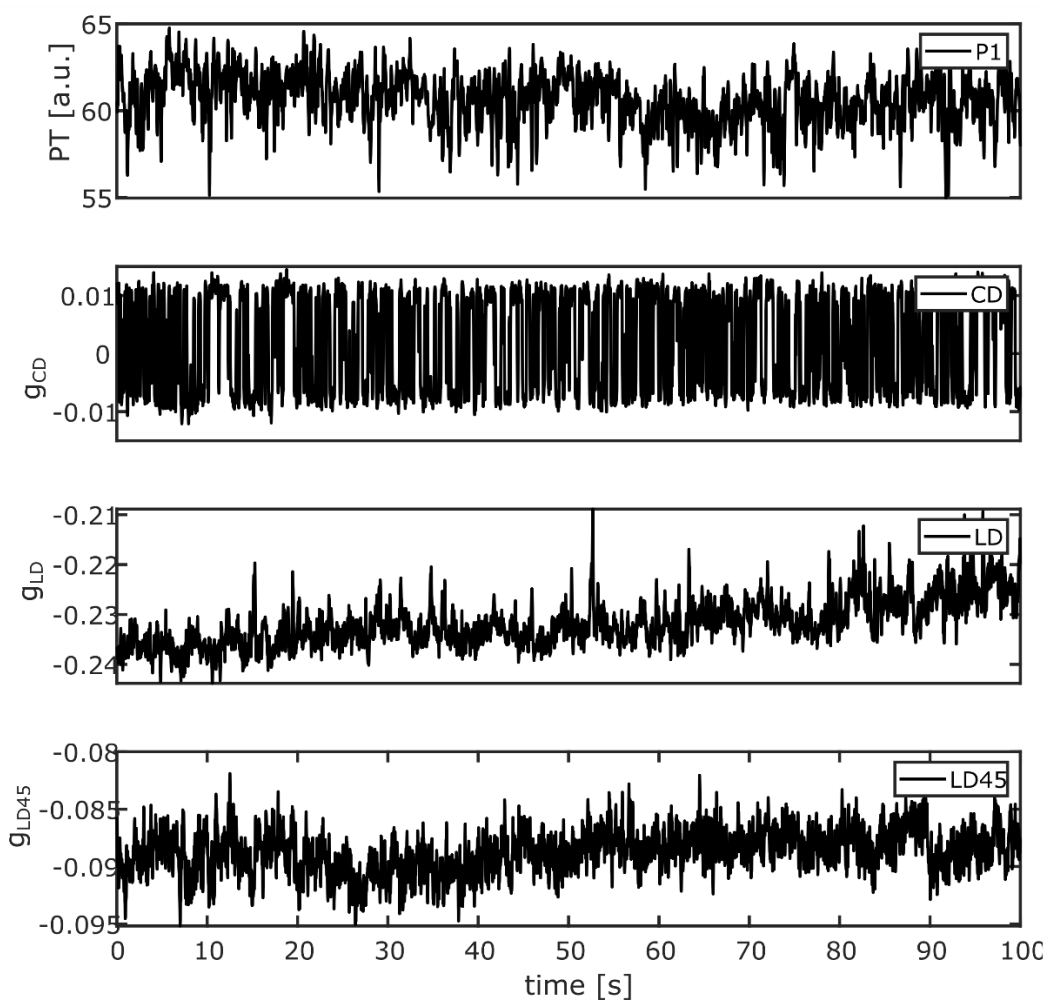

*Figure S13: Time traces of PT, CD, LD and LD45 of particle P1 over 100 s. The slight decrease in LD signal at the end of the time trace is due to slight defocusing. We see signal fluctuation between positive and negative values only for the CD time trace.*

## 7. Threshold analysis of field-dependent switching events for particle P1

Magnetization time traces of particle P1 at different magnetic fields presented in Fig.2 are reproduced here with a change-point analysis using the mid-point between the two histogram peaks as threshold (Figure S14). The threshold defines the two different states to calculate the residence times in the two states.

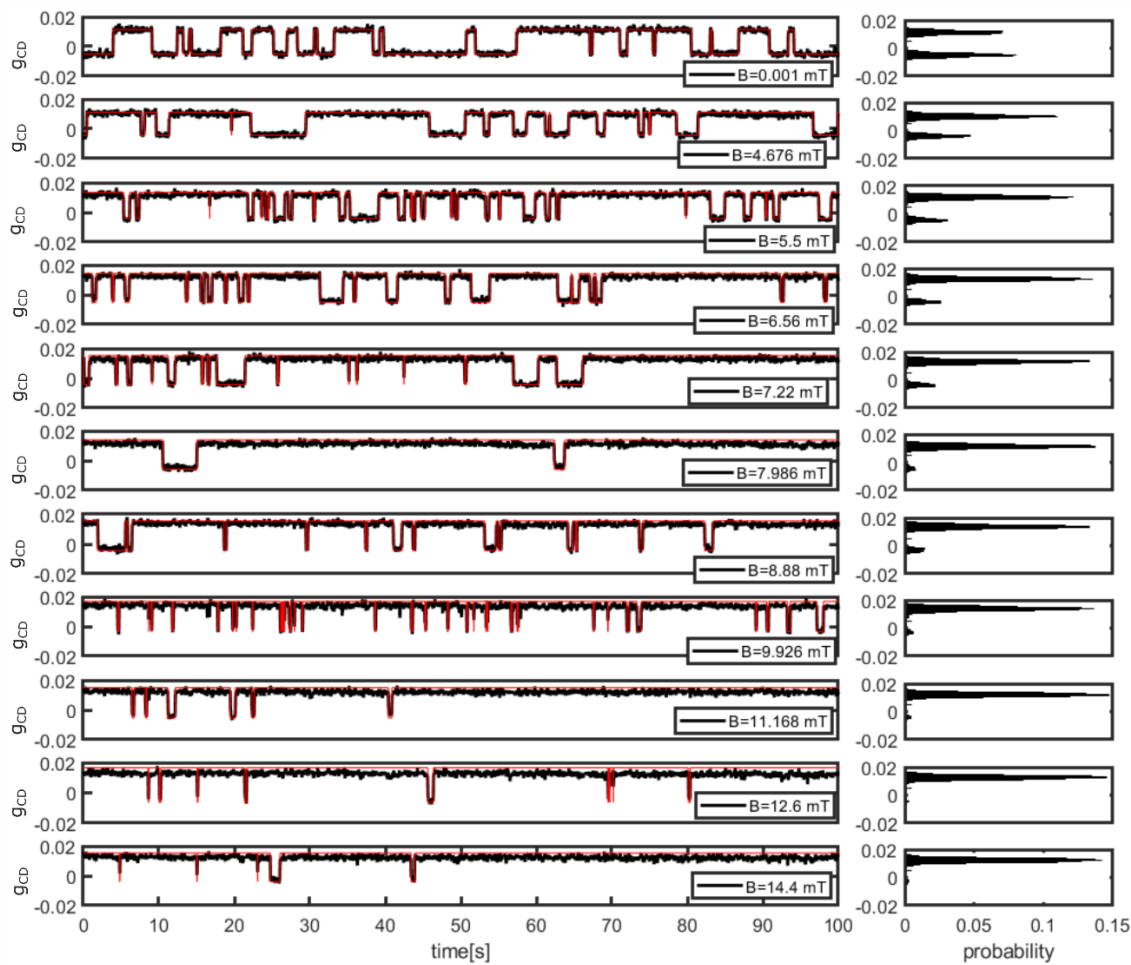

*Figure S14: External magnetic-field-dependent time traces of Fig. 2 of the main text. The threshold is defined to distinguish two states. The fits based on this threshold analysis are shown with solid red lines. The histogram of each time trace, providing the occupation probability of each state, is shown on the right. The dwell times in each state are deduced from the change-point analysis.*

## 8. Field-dependent time traces of particles P7 and P8

In the main text, we have shown field-dependent time traces of particle P1 only. Here, we show similar time traces for two more particles, P7 and P8 in Figure S15.

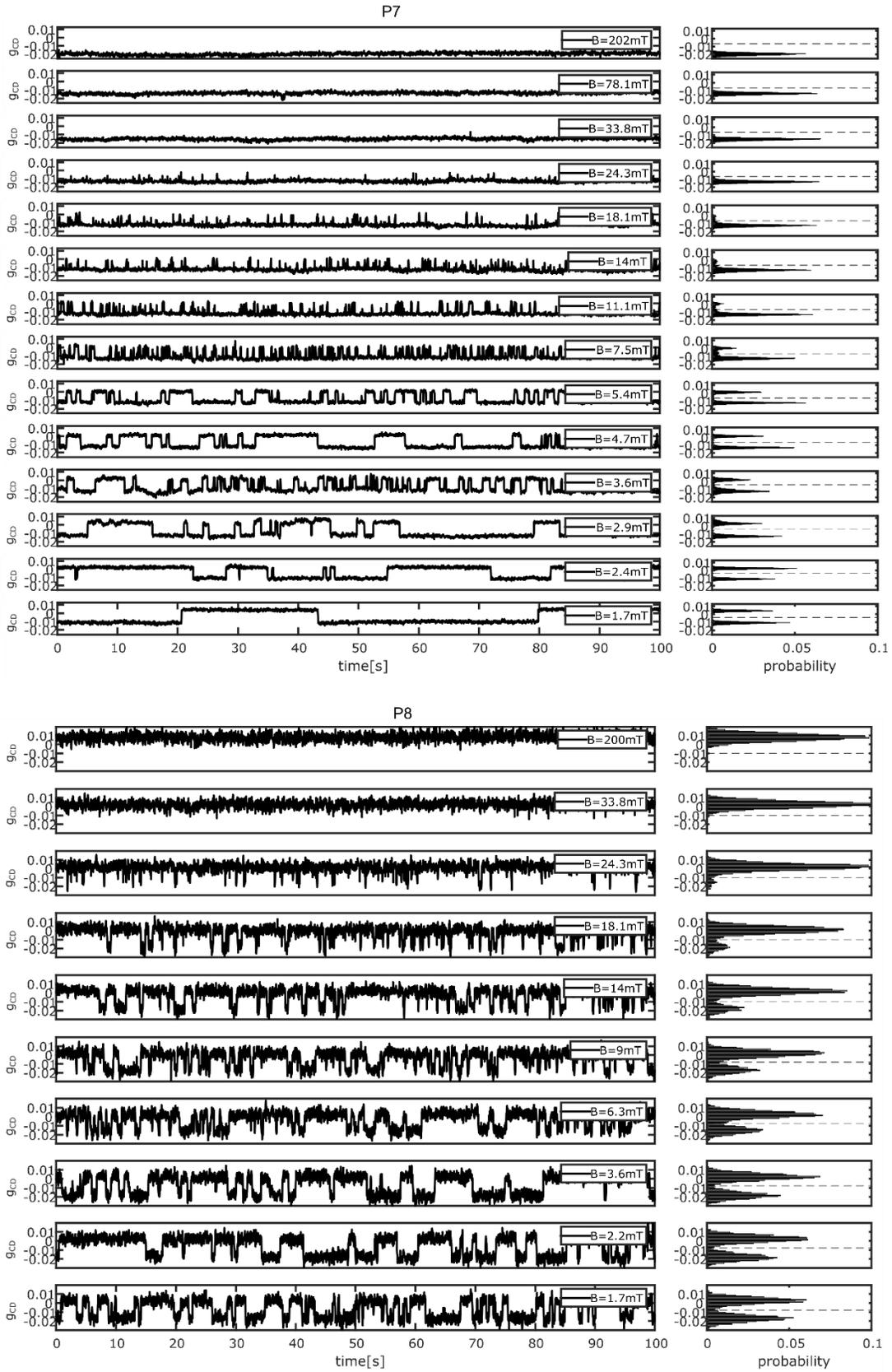

*Figure S15: Magnetic-field-dependent time traces of particles P7 and P8.*

The ratio of residence times in the two opposite magnetization states of particles P7 and P8 is presented in Figure S16 as a function of the applied magnetic field. For all values of the angle $\psi$, the Stoner-Wohlfarth model gives a very close-to-linear variation with a slope of $2\mu \cos\psi/k_B T$. Unfortunately, we did not measure the magnetization curves of these two particles and therefore we do not know the value of $\psi$. As can be seen from the time traces of these two particles, the field required to dominantly populate only one state has a value close to the one found for particle P1 in the main text. Therefore, we assumed $\psi$ to be close to 70° for particles P7 and P8. Considering $\psi=70°$, the magnetic moments for the P7 and P8 are about $3 \times 10^5$ Bohr magnetons and $0.95 \times 10^5$ Bohr magnetons. The magnetic moments of P7 and P1 are quite similar, in good agreement with their estimated sizes (23.4 nm diameter for P7 and 24.7 nm for P1). The magnetic moment of particle P8 is about 3 times lower than that of particle P7 which is close to the expected volume ratio, about 4.7. The slight mismatch could be explained by the thickness of their dead layers. As the surface-to-volume ratio is higher for smaller particles, the similar dead-layer thickness would lead to a reduced magnetic moment for smaller particles. The mismatch could also be due to different $\psi$ values as we do not know the exact $\psi$ values for these two particles.

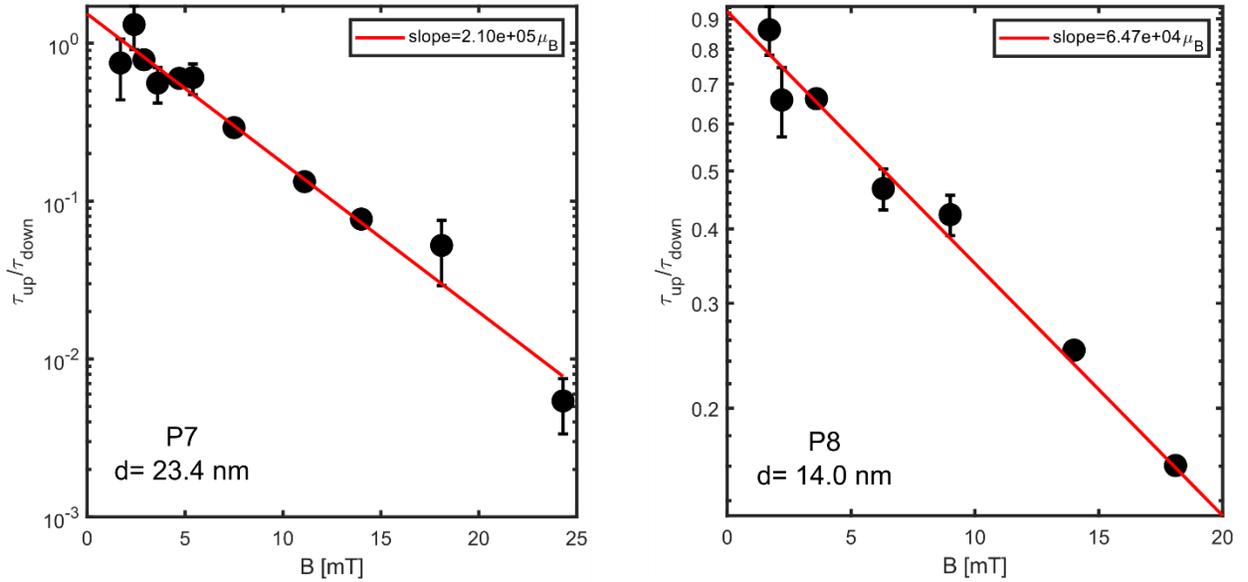

*Figure S16: External-field-dependent Boltzmann behavior fitted with the Stoner-Wohlfarth model for particles P7 and P8. The slopes extracted from the fits are $2.1\times10^5$ and $6.5\times10^4$ Bohr magnetons, respectively. The sizes of the particles are also mentioned in the insets. The error bars are the deviations from the fits.*

## 9. Temperature estimation

Simulations of scattering and heat transfer were carried out to calculate the absorbed power and the temperature variation of the nanoparticle, respectively. The nanoparticle scattering model is shown in Figure S17(a). Considering the symmetry of this model, we only take a quarter of the whole model to reduce the calculation time. The particle is at the center of the simulated volume and surrounded by the refractive medium. The mismatch of refractive index between substrate and medium has been ignored. Plane circularly polarized heating and probe beams are taken as background fields. The focusing configurations are different for these two beams, the focus beam waist of the heating beam is 1500 nm and that of the probe beam is 328 nm (0.61 $\lambda$/NA). The heating and probe powers absorbed by the particle can be obtained by a far-field scattering calculation. To get the temperature of the nanoparticle, we calculate heat diffusion from the nanoparticle to the substrate and medium. Here, we considered the different thermal conductivities of glass and hexadecane. The model and boundary conditions are shown in Figure S17(b). The spherical particle is located on the surface of the glass and surrounded by the liquid. Different heat conductivities have been assigned to the glass and to the liquid medium. The particle is set as a point heat source with the power obtained from the scattering simulation. The outside of this model is maintained at ambient temperature (293.15 K). The optical and thermal parameters of materials are shown in Table 1.

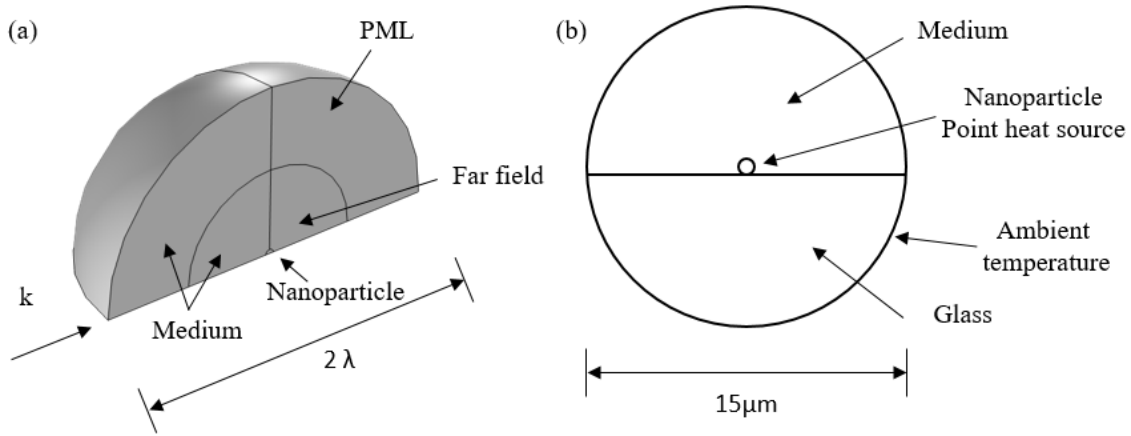

*Figure S17: COMSOL model used for the temperature calculation. See details in the text.*

|  | $k$ (W/(m·K)) | $C_p$ (J/(kg·K)) | $\rho$ (kg/m$^3$) | $n$ |
|---|---|---|---|---|
| magnetite | 6 | 890 | 5180 | $2.13 + i \times 0.86$ [3] |
| glass | 1.114 | 860 | 2510 | 1.516 |
| Oil | 0.14 | 1972.5 | 870 | 1.518 |
| hexadecane | 0.14 | 2215.1 | 770 | 1.433 |

*Table 1: Parameters used for the COMSOL simulation to calculate the temperature for magnetite nanoparticles in two photothermal media, oil and hexadecane.*

## 10. Temperature dependence of the switching events of particle P1

As we vary the particle temperature through variable heating and probe powers, we need the absorption coefficients of magnetite at these two wavelengths to estimate the particle temperature. A number of literature data can be found for the refractive indices of magnetite. Here we selected two values from the works of Huffman et al.[4] and that of Triaud et al.[3] Using the COMSOL model described above, we calculated the temperatures and deduced the residence times in up- and down-states shown in Figure S18. Arrhenius fits to both of these temperature dependences show a similar anisotropy energy barrier. In the main text, we chose the refractive index database of Triaud et al.

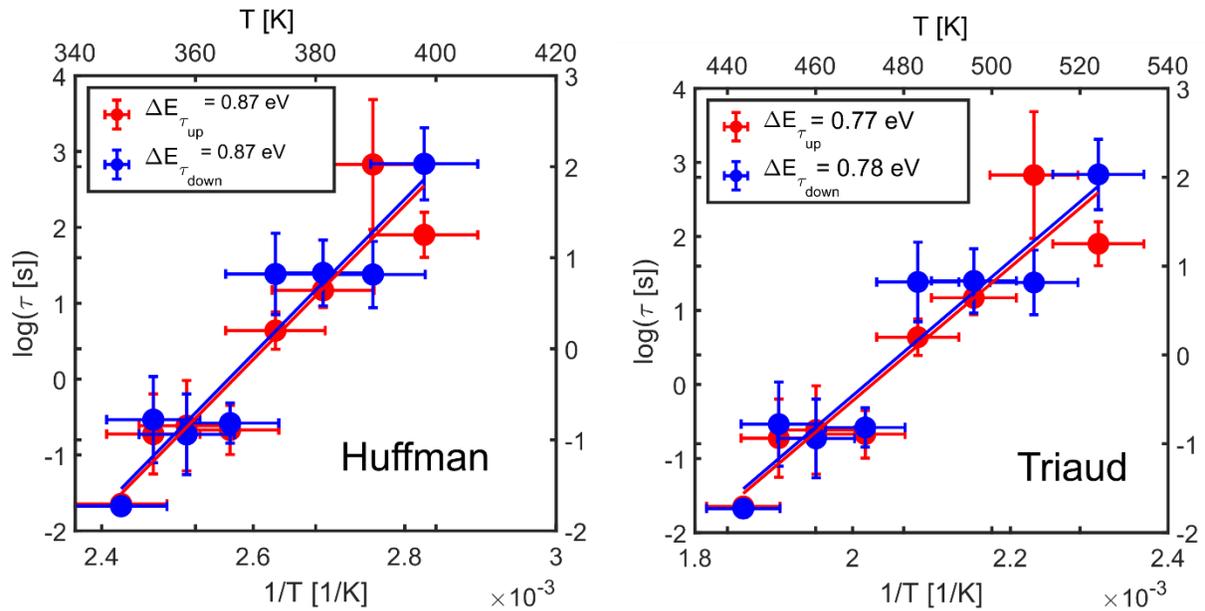

*Figure S18: Temperature dependence of residence times in the two magnetization states of particle P1 fitted with the Stoner-Wohlfarth model. The temperature is calculated on the basis of two literature values of the optical absorption of magnetite, by Huffman[4] and Triaud.[3] Both of these temperature dependences yield similar values for the anisotropy energy barrier.*

## 11. Dynamical heterogeneity of a particle measured over several hours

To improve the characterization of the dynamical heterogeneity of magnetic switching, we followed a single magnetite nanoparticle (particle A1) for several hours, but not continuously. We allowed for several waiting times without laser exposure. The results are shown in Figure S19. The particle shows various magnetic behaviors over this longer time period. The mechanism behind the transitions between these behaviors is not well understood. We speculate that composition fluctuations due to possibly reversible (photo-)oxidation events and/or structural or surface defects and/or PVP ligands on the surface may influence the topography of the magnetic energy landscape, as seen for example in the simulations by Winklhofer et al.[5]

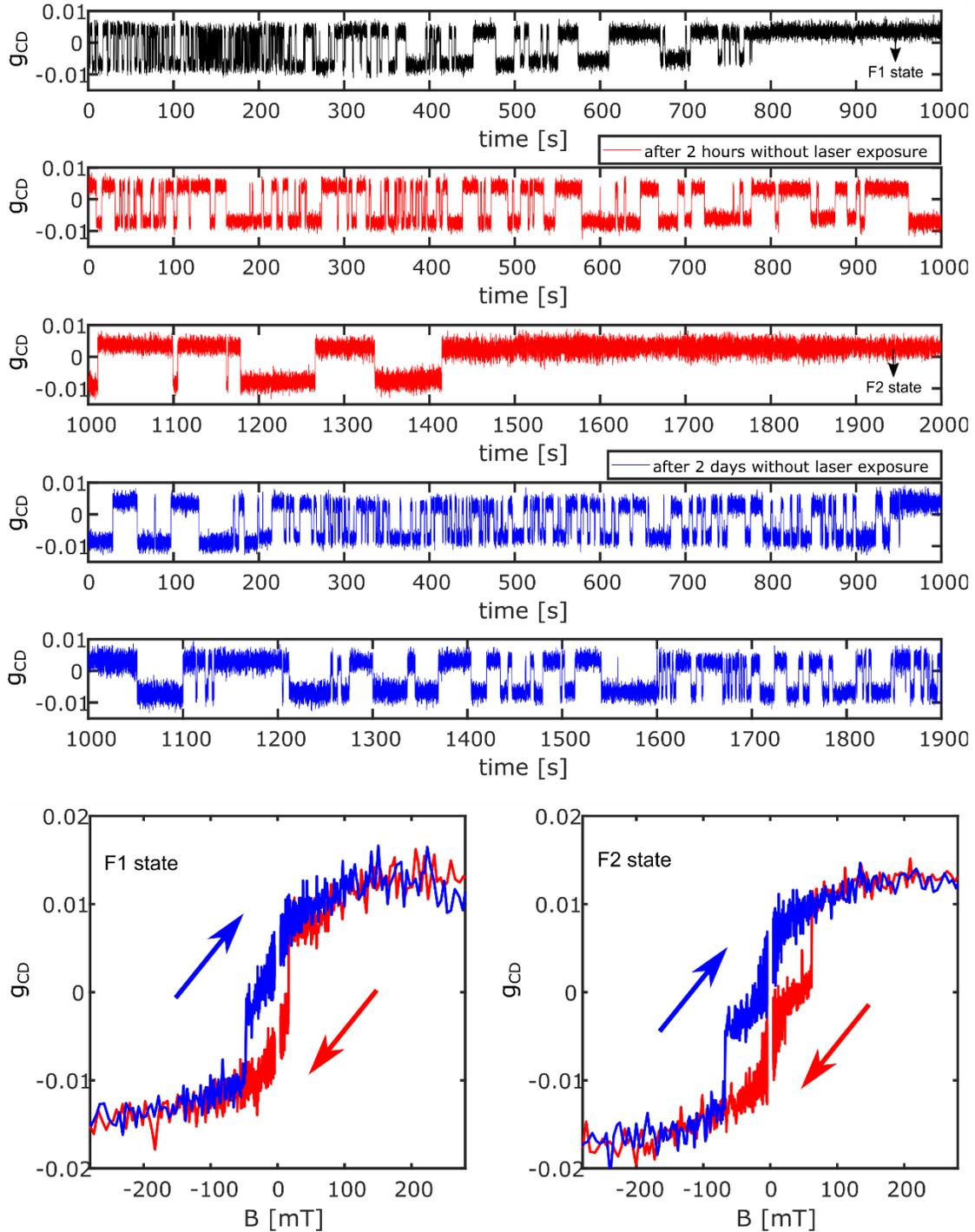

*Figure S19: Dynamical heterogeneity in the switching behavior of a single particle (A1) measured over several hours. The black time trace shows dynamical heterogeneity over a continuous measurement of 1000 s. The intermediate switching behavior slowly becomes ferromagnetic (F1 state) after 800 s, as evidenced by the hysteresis of the magnetization curve measured in the F1 state (bottom left). After 2 hours without laser exposure, the particle again resumes switching, as shown by the red time trace. Over the next period of 2000 s, the particle reverts to the ferromagnetic (F2) state, as evidenced by the hysteresis loop in the magnetization*

*curve measurement of the F2 state (bottom right). After 2 days without laser exposure, the particle resumes switching again and remains in the intermediate switching state for over 1900 s. Arrows indicate the direction of magnetic field sweep.*

12. Dynamical heterogeneity of UV-plasma-cleaned particles

To check if the surface ligands alone are responsible for the dynamical heterogeneity, we cleaned the magnetite nanoparticles in UV-plasma. Switching measurements of such particles (particle B1 and particle B2) are shown in Figure S20. These particles still show pronounced dynamical heterogeneity. Therefore, we conclude that surface ligands cannot be the only cause of dynamical heterogeneity.

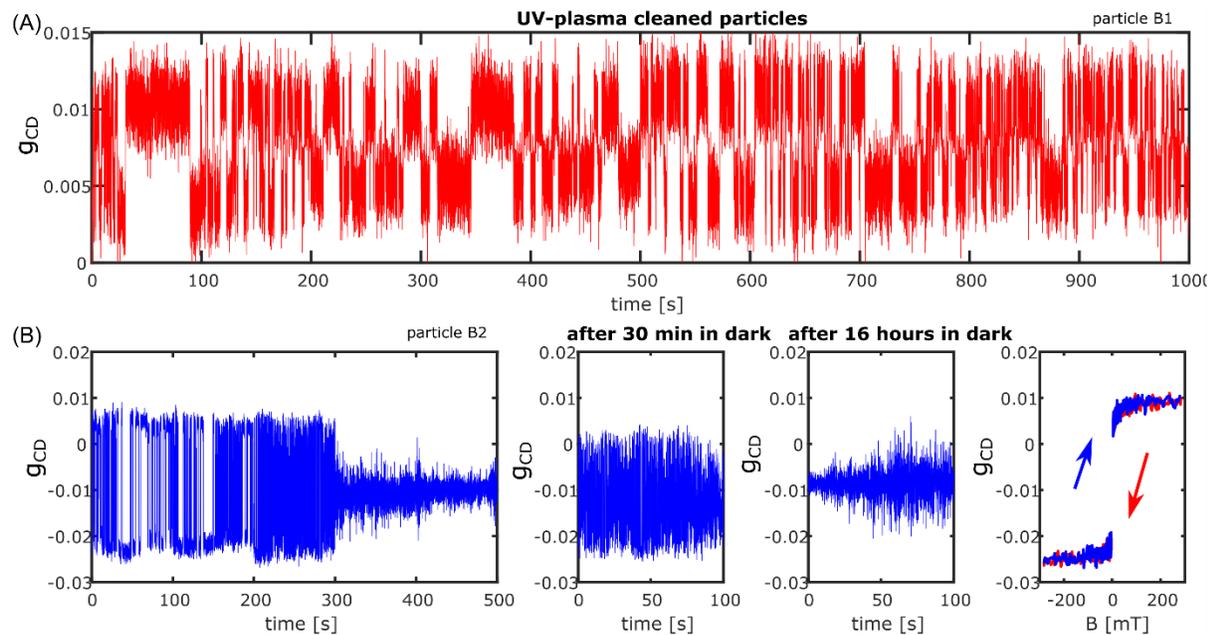

*Figure S20: Dynamical heterogeneity in the switching behavior of two particles (particle B1 and particle B2) which were UV-plasma cleaned. (A): Particle B1 shows a change of switching rates over time but still remains in the intermediate magnetic state. (B): Particle B2 shows a change from the intermediate switching magnetic behavior to a quickly switching, superparamagnetic state. Without laser exposure for 30 min, the particle seems to slow somewhat in its switching rate, but still remains very fast compared to the earlier intermediate state observed at the beginning of the trace. Here, 16 hours without laser exposure did not change much the superparamagnetic behavior, as evidenced by the hysteresis-free magnetization curve. Arrows indicate the direction of magnetic field sweep.*

13. Dynamical heterogeneity of a particle covered with HfO$_2$

To check whether surface oxidation influences the dynamical heterogeneity behavior of our magnetite nanoparticles, we deposited a thin (5 nm) layer of HfO$_2$ by ALD (atomic layer deposition) which would protect the sample from oxidation. We measured two particles (particle C1 and particle C2) as shown in Figure S21. Both these particles show changes of switching rates over time, and in both cases, the switching rate becomes faster over time. In the case of particle C2, we observed three states at the beginning of the trace, which later changed to two states, upon disappearance of one of the states, another possible manifestation of dynamical heterogeneity. The appearance of three states in particle C2 is not understood, and is possibly due to crystalline anisotropy and shape anisotropy both influencing the magnetic energy landscape of particle C2.

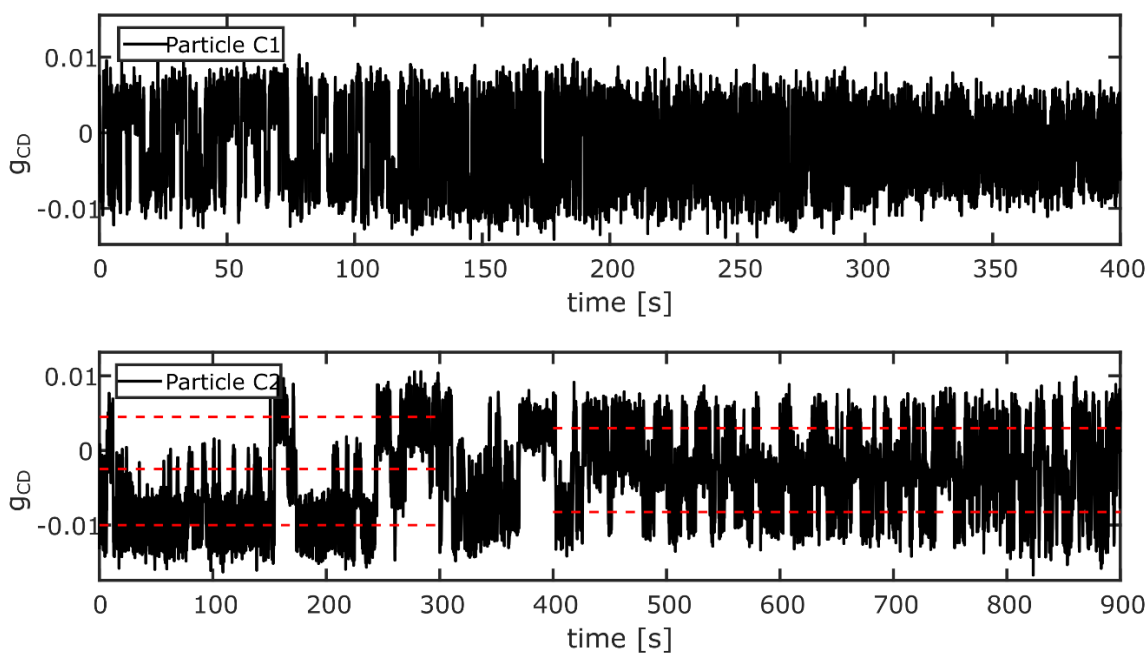

*Figure S21: Dynamical heterogeneity of the switching of two particles (particle C1 and particle C2) measured in a magnetite nanoparticle sample coated by about 5 nm of HfO$_2$. Both these particles show changes of switching rates over time i.e., dynamical heterogeneity.*

14. Dynamical heterogeneity of a particle which was plasma-cleaned and covered with HfO$_2$

To avoid the effects of both surface oxidation and ligands on the dynamical heterogeneity, we prepared a sample which was plasma-cleaned and subsequently covered with 5 nm HfO$_2$. We measured time traces of three particles (particle D1, particle D2 and particle D3) as shown in Figure S22. Although particle D2 shows weak dynamical heterogeneity, the other two particles show quite pronounced dynamical heterogeneity. This result suggests that the changes which induce dynamical heterogeneity are probably taking place inside the particle. Internal charge reordering events upon heating have previously been reported.[6]

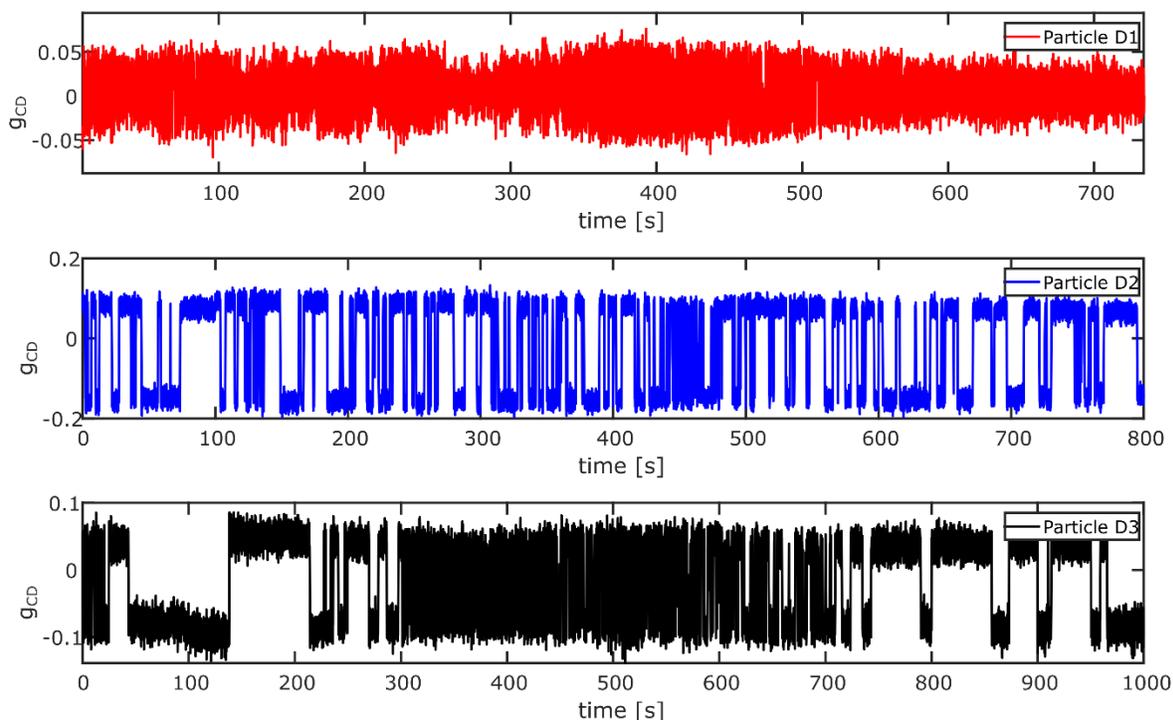

*Figure S22: Time traces of three particles (particle D1, particle D2 and particle D3) measured in a plasma-cleaned sample covered with 5 nm HfO$_2$. Time traces showing the dynamical heterogeneity behavior.*

15. Discussion of the terminology of the g-factor and the dissymmetry factor g$_{CD}$

In the field of single-particle circular dichroism spectroscopy, the so-called dissymmetry factor i.e., the differential absorption of left and right circularly polarized light (i.e., CD) normalized by total absorption is traditionally called "g-factor". To avoid confusion with the Landé factor g in magnetism, well-known in the calculations of magnetic moments and angular momentum states of atoms, we referred the CD g-factor as "g$_{CD}$". We recommend the use of this terminology in future work.

16. Estimation of inter-particle magnetic field in our sample

When particles are close-by, the dipolar field of nearby particles can influence a particle's magnetic properties. If the radius of the particle is R and the distance of the particle from the near-by particle is r, the mean field produced by the near-by particle is $(R/r)^3 * B_s$, where $B_s$ is the field at the particle surface. For our case, for example particle P1 in the main text (see Figs. 1E and 2A), $B_s$ is above 10 mT, $R$ is about 20 nm and $r$ is more than 1 µm. Therefore, the mean field produced by a particle in a distance of 1 µm on the particle P1 would be about 80 nT. Due to such a low value, we neglected the mean field created by other particles in the magnetization switching analysis.

17. Magnetic field vs distance

Fig. S20 shows the magnetic field measured with a Gaussmeter upon varying the distance between the Hall probe and the magnet. A typical field required to saturate a particle like P1 in the main text is below 20 mT at which value the external magnet is more than one cm away from the sample. At such a distance, we assumed the magnetic field to be uniform. However, the magnetic field near the sample could be more heterogeneous, which could introduce some inaccuracy on the largest field values.

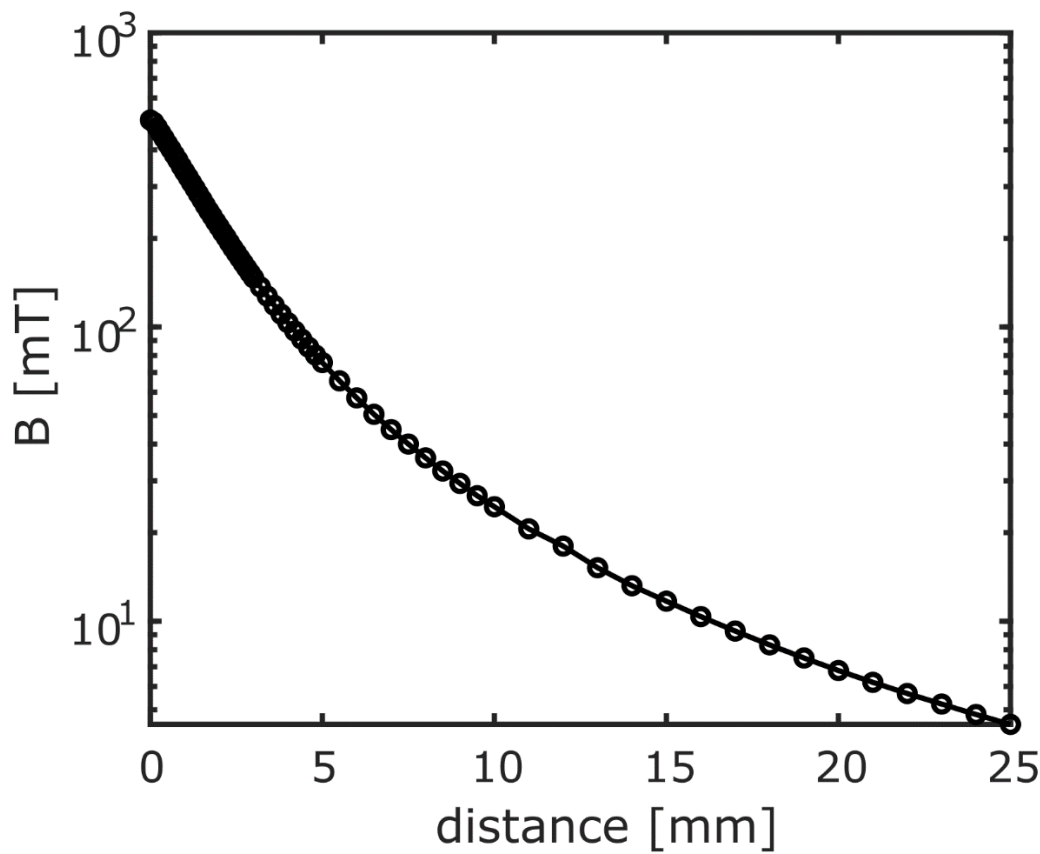

*Figure S23: Plot of the magnetic field measured with a Gaussmeter vs the distance between the magnet and the Hall probe.*

18. Basic principle of the PT MCD technique

Details about the PT MCD method can be found in our previous paper[7]. Here we briefly recall the basic principles of the method. The PT MCD method is based on photothermal microscopy. When an absorbing nanoobject is illuminated with light (which we call the heating beam), the nanoparticle creates heat through non-radiative relaxation and dissipates it into its surroundings. The heat dissipation creates a temperature profile (which is an *1/r* profile in steady state; *r* is the distance from the particle). Due to the thermo-refractive behavior of the surrounding medium, the temperature profile creates a refractive index profile which is called the thermal lens. A second beam (which we call the probe beam) is used to probe the thermal lens. The scattered probe light interferes with the incident probe beam (either reflected or transmitted). A modulated interference signal due to modulation of the heating laser is filtered using a sensitive lock-in amplifier, which provides the photothermal signal. Therefore, the photothermal signal gives information about the absorption of the nanoobject. This basic principle provides the contrast of photothermal microscopy.

Circular dichroism (CD) is defined as the differential absorption of left and right circularly polarized light. It is a specific property of chiral nanoobjects. When a chiral nanoparticle is heated by absorption of a beam with polarization modulated between left- and right-circular states, the chiral nanoparticle produces a modulated photothermal signal which is called photothermal circular dichroism (PT CD).

Photothermal magnetic circular dichroism (PT MCD) is the differential absorption signal of a magnetic nanoparticle in the presence of an external magnetic field. It is due to the polar magneto-optical Kerr (MOKE) effect. The magnetic field is applied along the optical axis. With the flip of magnetic field direction, the magnetic moment orients along the magnetic field and thus a signature of PT MCD is the flip of the sign with flip of the field direction. The strength of the PT MCD signal is proportional to the magnetic moment of the nanoparticle and the so-called $g_{CD}$ factor is proportional to the magnetization of the particle.

19. Detection sensitivity

The particle P1 in the main text has a magnetic moment of $4\times10^5$ Bohr magnetons. As shown in Fig. 1C, the signal-to-background of MCD signal is more than 10 with an integration time of 100 ms. Therefore, the detection sensitivity of our method on that time scale is better than $4\times10^4$ Bohr magnetons. As an example, the particle P8 shown in Figs. S12-13 has a magnetic moment of $0.95\times10^5$ Bohr magnetons which is about 2.5 times the detection sensitivity. We have measured the magnetization switching of this particle and the magnetic moment is calculated from the fit according to the Stoner-Wohlfarth model. The detection sensitivity can still be improved with an increase in integration time.

## 20. Possible applications of PT MCD

PT MCD can be applied to study magnetization switching under the influence of applied external perturbations such as electric fields, microwave or optical pulses. As a relatively large sample area of many single nanoparticles can be influenced simultaneously by those perturbations, extended statistics of single-particle switching could be established in a relatively short time with a high throughput. More specifically, the switching mechanism could be investigated in each of these cases, either by comparing several repeated switching events of the same particle, or by observing switching events of several different particles and correlating them to their individual properties such as magnetization anisotropy and easy-axis orientation. The influence of external parameters (temperature, electric and magnetic fields, chemical reactions, etc.) could be mapped on a single-particle basis and correlated with other properties, such as is currently done for single biomolecules.

PT-MCD can also be applied to many magnetic nanomaterials, such as anti-ferromagnetic nanoplatelets[8] and metal-insulator-metal devices[9]. Single-particle studies will remove ensemble-averaging and expose correlations between particle structure and their magnetic properties. Such a study can help us to investigate domain wall propagation and nucleation in magnetic materials on the single-particle level.

## 21. Calculation of the critical radius for a single-domain magnetite particle

We calculated the critical radius ($r_{critical}$) for a single domain magnetite nanoparticle using the following formula

$$r_{critical} = \frac{36\sqrt{AK}}{\mu_0 M_s^2}$$

where, $A = 10^{-11}$ J/m, $K = 10^4$ J/m$^3$, $\mu_0 = 1.25 \times 10^{-6}$ N/A$^2$ and $M_s = 446 \times 10^3$ A/m. The calculated critical radius is about 45 nm which is in good agreement with the data available from the literature[10].

## 22. Time traces of MCD signals for particles P1 to P6 as labelled in Fig. 1 in the main text

Figure S21 shows time traces of the MCD signal of the six particles labelled in the main text (Fig. 1) P1 to P6. The time traces indicate that particle P1 behaves differently from the other particles. Particle P1 shows intermediate magnetization behavior that displays switching of the MCD signal between two values.

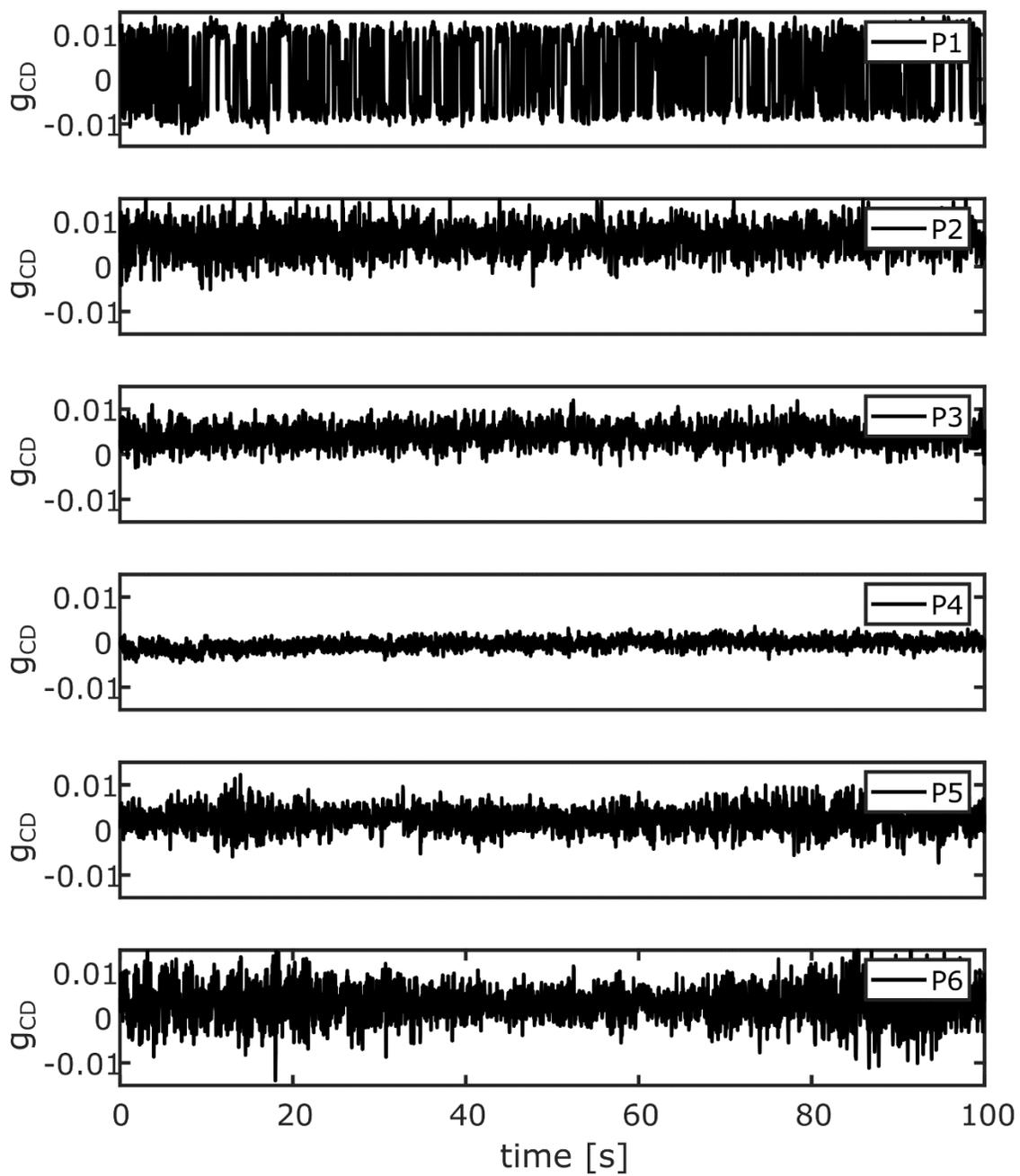

*Figure S24: Time traces of MCD signal over 100 s for particles P1-6 from Fig. 1 in the main text.*

## 23. Attempt frequency obtained from the Arrhenius fit of temperature dependent switching of Particle P1 in the main text

According to Néel-Brown theory, the lifetime of the up or down state of magnetization switching follows an Arrhenius dependence,

$$\tau = \frac{1}{f_0} exp\left(\frac{\Delta E}{k_B T}\right)$$

where $\tau$ is the lifetime of a magnetization state, $f_0$ is the attempt frequency, $\Delta E$ is the energy barrier, $k_B$ is the Boltzmann constant and $T$ is the absolute temperature.

In most studies, the attempt frequency[11] is considered as a constant factor in the order of $10^{10}$ Hz[12]. Krause et al. showed that the attempt frequency depends on the morphology of the particles[13]. In our case, we obtained the attempt frequency from the Arrhenius fit as shown in Fig. 2c in the main text and its value is about $10^8$ Hz, which is close to the reported value in the reference[14]. It would be interesting to study size- and shape-dependent magnetization switching using the PT MCD technique; however, this is beyond the scope of the current study.

24. Magnetization curve of a ferromagnetic particle

Fig. S22 shows magnetization curves of a ferromagnetic particle calculated using the Stoner-Wohlfarth model (detailed discussion in the section "Stoner-Wohlfarth model"). The magnetization curve is calculated at different angles of the easy axis with respect to the direction of the applied magnetic field. The magnetization curve with the easy axis angle of 90° i.e., the particle's easy axis along the sample plane, looks similar to the magnetization curve of a superparamagnetic particle with similar easy axis orientation (see Figure S4). Therefore, it is difficult to distinguish between superparamagnetic and ferromagnetic behavior from the magnetization curve when the particle's easy axis is perpendicular to the applied magnetic field.

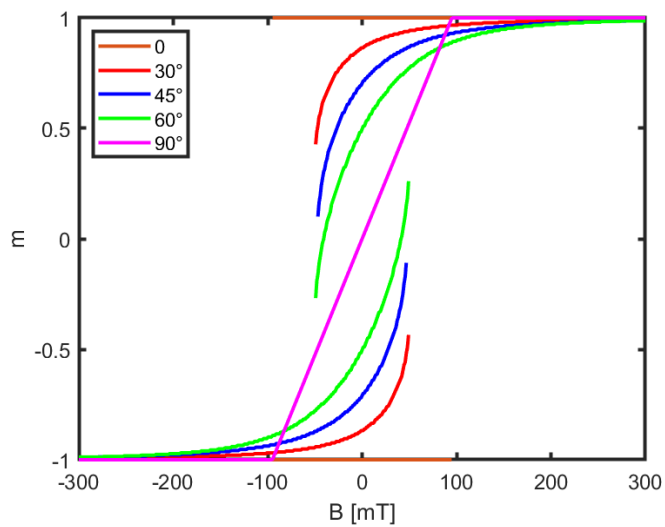

*Figure S25: Magnetization curve of a ferromagnetic particle (aspect ratio of 1.5) calculated using the Stoner-Wohlfarth model. The angles mentioned in the inset are the angle of the easy axis with respect to the direction of the external field.*

25. Analysis of dynamical heterogeneity for another time trace as in Fig. 3

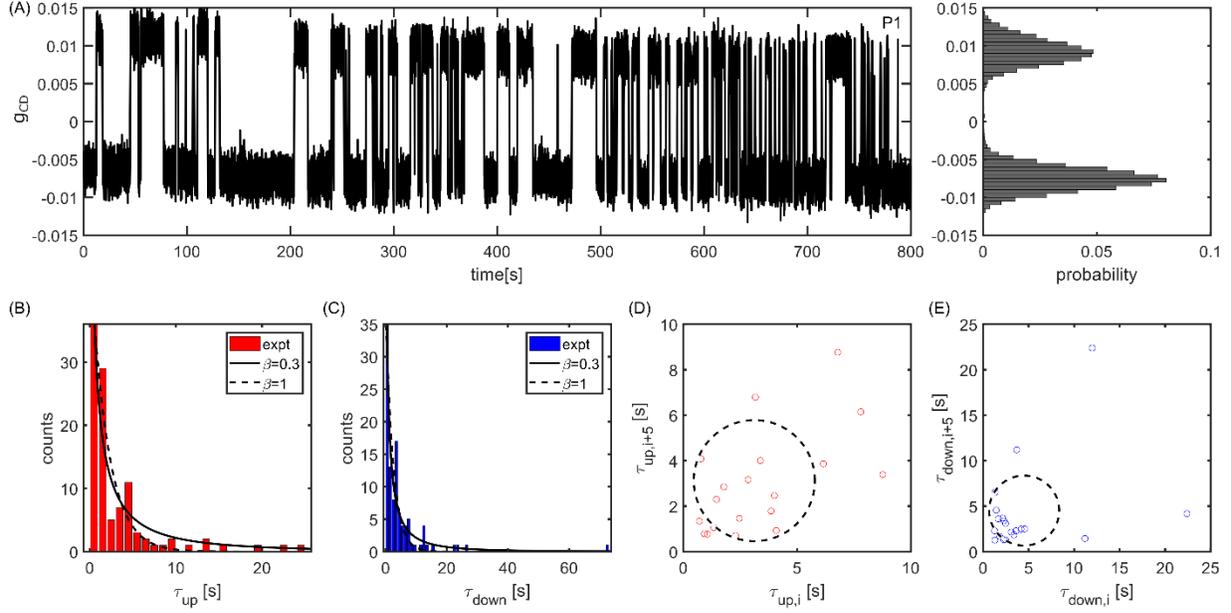

*Fig. S23: Dynamical heterogeneity of magnetization switching of particle P1. A: Time trace of magnetization switching over 800 s. The corresponding histogram of $g_{CD}$ is shown on the right. B, C: Histograms of $\tau_{up}$ and $\tau_{down}$ with stretched-exponential fits (stretching exponents $\beta$ given in insets). D, E: Correlation plots of successive averages of $\tau_{up}$ and $\tau_{down}$, averaged over five successive events. Strong fluctuations of these averages lead to correlation points out of the dashed discs (see details in the main text).*

26. Size distribution of single magnetite nanoparticles

The photothermal signal of a single small particle is proportional to its volume and to its Clausius-Mossotti factor. With a particle of a known size and material, one can calibrate the photothermal signal to determine the size of a particle of a different material. We used 10 nm gold nanoparticles to calibrate the photothermal signal of magnetite nanoparticles to obtain their sizes, with the known refractive indices of magnetite. We found two sets of data for the magnetite refractive index, as mentioned earlier in the text. For both data sets, the size distribution obtained by the above-mentioned calibration matches the size distribution obtained from the TEM measurement reasonably well. However, the data of Triaud et al. [3] give a better agreement than those of Huffman et al. [4], as shown in Figure S24.

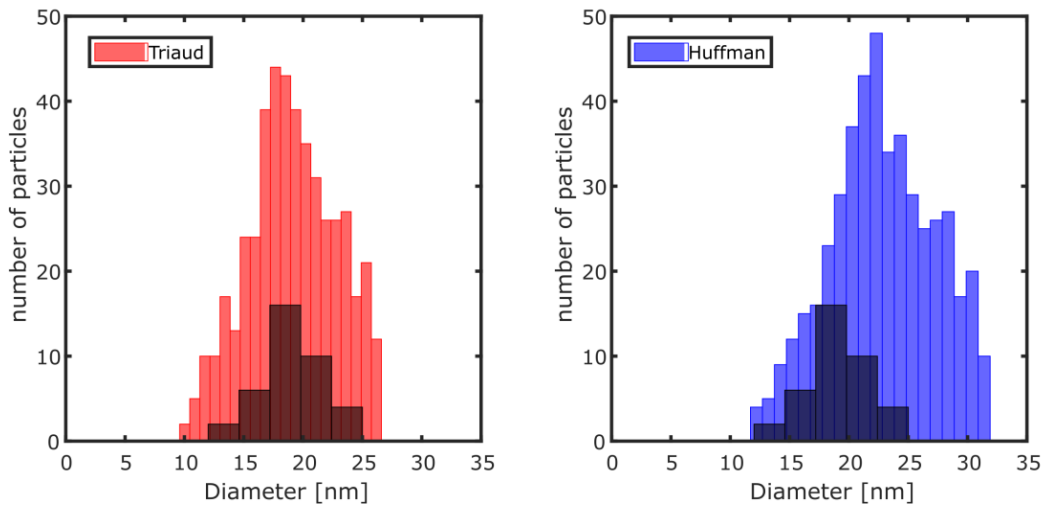

*Figure S24: The size distribution of single magnetite nanoparticles obtained from their photothermal signals as shown in Figure S1, using two refractive index databases; (a) Triaud and (b) Huffman as mentioned earlier in the text. The size distributions are compared to the size distribution obtained from TEM as shown in Figure S2.*

27. Absorption spectra of magnetite:

Figure S25 shows the absorption spectra of 20 nm magnetite nanoparticles according to the manufacturer (Nanocomposix).

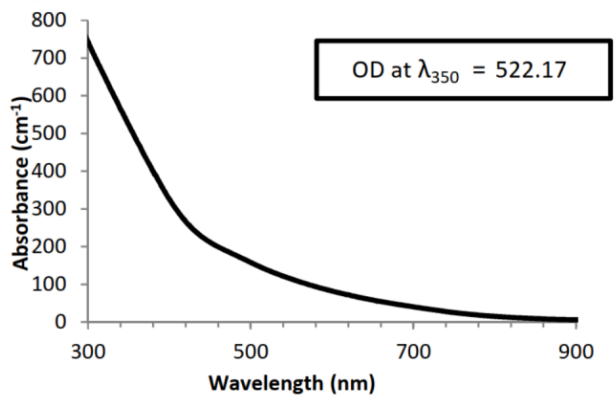

*Figure S25: Absorption spectra of magnetite provided by the manufacturer.*